\documentclass[12pt]{article}

\usepackage{amsfonts,amsmath}
\usepackage{latexsym}

\usepackage{epsfig}
\usepackage{psfrag}

\usepackage{epsfig}
\def\void{}
\def\labelmark{\marginpar{\small\labelname}}

\newenvironment{formula}[1]{\def\labelname{#1}
\ifx\void\labelname\def\junk{\begin{displaymath}}
\else\def\junk{\begin{equation}\label{\labelname}}\fi\junk}%
{\ifx\void\labelname\def\junk{\end{displaymath}}
\else\def\junk{\end{equation}}\fi\junk\labelmark\def\labelname{}}

{\ifx\void\labelname\def\junk{\end{array}\end{displaymath}}
\else\def\junk{\end{array}\right.\end{equation}}
\fi\junk\labelmark\def\labelname{}\def\junk{}
}

\newcommand{\beq}{\begin{formula}}
\newcommand{\eeq}{\end{formula}}
\newcommand{\beqv}{\begin{formula}{}}

\newcommand{\oh}{\frac{1}{2}}

\newcommand{\bea}{\begin{eqnarray}}
\newcommand{\eea}{\end{eqnarray}}
\newcommand{\beas}{\begin{eqnarray*}}
\newcommand{\eeas}{\end{eqnarray*}}
\newcommand{\beqs}{\begin{displaymath}}
\newcommand{\eeqs}{\end{displaymath}}

\newcommand{\br}{\langle}
\newcommand{\kt}{\rangle}

\newcommand{\cS}{{\cal S}}
\newcommand{\cD}{{\cal D}}
\newcommand{\cC}{{\cal C}}

\newcommand{\ben}{\begin{equation}}
\newcommand{\een}{\end{equation}}

\newcommand{\bdm}{\begin{displaymath}}
\newcommand{\edm}{\end{displaymath}}

\newcommand{\pa}{\partial}
\newcommand{\D}{{\cal D}}

 \newtheorem{theorem}{Theorem}
 
\newtheorem{lemma}{Lemma}

\newtheorem{definition}{Definition}

\newtheorem{proposition}{Proposition}

\newtheorem{remark}{Remark}

\begin{document}

\hfill

\addtolength{\baselineskip}{0.5\baselineskip}
 \topmargin 0pt
 \oddsidemargin 5mm
 \headheight 0pt
 \topskip 0mm


\begin{center}

{\Large \bf The structure of spatial slices of 3-dimensional causal triangulations}

\medskip
\vspace{1 truecm} 



{\bf Bergfinnur Durhuus}\footnote{email: durhuus@math.ku.dk}

\vspace{0.4 truecm}

Department of Mathematical Sciences

University of Copenhagen, Universitetsparken 5

DK-2100 Copenhagen \O, Denmark

 \vspace{1.3 truecm}

{\bf Thordur Jonsson}\footnote{email: thjons@hi.is}

\vspace{0.4 truecm}

Division of Mathematics, The Science Institute

University of Iceland, Dunhaga 3

IS-107 Reykjavik, Iceland

\end{center}
 \pagestyle{empty}

\hfill

\noindent {\bf Abstract.}  We consider causal 3-dimensional triangulations with the topology of $S^2\times [0,1]$ or 
$D^2\times [0,1]$ where $S^2$ and $D^2$ are the 2-dimensional sphere and disc, respectively.
These triangulations consist of slices
and we show that these slices can be mapped bijectively onto a set of certain coloured 2-dimensional cell complexes
satisfying simple conditions.  The cell complexes arise as the cross section of the individual slices.

\newpage

 \pagestyle{plain}

\section{Introduction}  
We investigate in this paper a class of problems that arise in 
the dynamical triangulation approach to 3-dimensional gravity restricted to the case of so-called causal triangulations. For an introduction to the dynamical triangulation approach to discrete quantum gravity we refer to \cite{book} and an account of causal dynamical triangulations can be found in \cite{al4}. For the case of 3-dimensional gravity in particular, one may consult \cite{adj1} and \cite{al2}. 

In \cite{DJ} we gave a proof that the number $N(V)$ of causal 3-dimensional triangulations homeomorphic to a 3-sphere and consisting of $V$ tetrahedra is exponentially bounded,
$$
N(V)\leq C^V\,,
$$
where $C$ is some positive constant. Validity of this bound is crucial in order for the relevant correlation functions
to exist, and thereby defining the discretised model (see \cite{book,al4}). The first step of the argument leading to this bound was to decompose the triangulations into slices (which is possible because of the causal structure defined below) and to show that it is sufficient to establish the bound for such causal slices. The number of triangulated causal slices 
was then shown to be bounded by the number of certain coloured 2-dimensional cell complexes homeomorphic to the 2-sphere.  These cell complexes arise 
as mid-sections of the slices. Finally, bounding the number of those 2-dimensional cell complexes can be done using well known techniques. 

The method of associating a coloured 2-\-dimensional cell complex with a causal triangulation has been applied earlier by other workers in the field \cite{al2,al3} and also used in numerical simulations and in combination with matrix model techniques to extract properties of the model. It has, however, not been established exactly what class of 2-dimensional complexes can occur as mid-sections of a causal slice. As noted in \cite{DJ}, see also \cite{al2} Appendix B, 
some non-trivial constraints have to be imposed on top of its homeomorphism class. It follows that the connection between the model defined in terms of triangulations and the one realised in terms of a specific class of coloured 2-dimensional cell complexes is obscure.  It might be the case that the precise class of cell complexes is unimportant in a possible scaling limit of the model, see \cite{al3}, but this remains to be investigated in detail. 

In this paper we provide a complete characterisation of the coloured 2-dimensional cell complexes that correspond to 
3-dimensional causal slices. In fact, we shall consider not only the standard notion of causal triangulations homeomorphic to $S^2\times [0,1]$, where $S^2$ denotes the 2-sphere, which in this paper will be referred to as causal sphere-triangulations, but find it useful to generalise the notion of causal triangulation to manifolds homeomorphic to $D^2\times [0,1]$, where $D^2$ is the 2-dimensional disc, and $\partial D^2\times [0,1]$ may be viewed as the time-like part of the boundary while the two discs $D^2\times \{0\}$ and $D^2\times \{1\}$ form the spatial parts of the boundary. Such triangulations will be called causal disc-triangulations and the corresponding slices will be called causal disc-slices. The coloured cell complexes, defined in Sec.\ 3, are in this case homeomorphic to $D^2$. 
The main result of this paper, Theorem 1, which is proved in Sec.\ 4, states that there is a bijection between the causal disc-slices and the coloured cell complexes.
In Sec.\ 5 we generalise this result to causal sphere-slices in Theorem 2.

  
These results might serve as a starting point for an exact enumeration of causal slices by applying well known techniques for planar surfaces or maps to the coloured cell complexes homeomorphic to the disc. This interesting combinatorial problem is more complicated than those previously considered because of the colouring and the constraints identified in this paper. The result might also be instrumental in finding an appropriate matrix model generating exactly the desired causal slices or as an aid in designing effective numerical algorithms for simulations. These issues are, however, beyond the scope of the present paper.

\section{Preliminaries and notation} 

 We will use notation consistent with that of \cite{DJ}. For the reader's convenience we briefly recall the main conventions, restricting the discussion to the 3-dimensional case. The basic building blocks of our triangulations are tetrahedra or 
 3-simplices whose vertices have one of two colours: red or blue. Generally we will denote an unoriented simplex with vertices $x_1,\dots, x_n$ by $(x_1\dots x_n)$.  If all the vertices in a simplex have the same colour we say that the simplex is monocoloured.
This means that if $x$ and $y$ are red vertices and $e=(xy)$ is a 1-simplex then we say that $e$ is red, and a triangle $\Delta=(xyz)$ is red if its vertices (or edges) are red, etc.  If a simplex is not monocoloured we say it is two-coloured. It is assumed that all tetrahedra are two-coloured. Thus the tetrahedra come in three types: type (3,1) with three red vertices, type (2,2) with two red vertices and type (1,3) with one red vertex.

We recall that an abstract simplicial complex $K$ is defined by its vertex set $K^0$, which is assumed to be finite, and a collection of subsets of 
$K^0$, called simplices, such that if $\sigma$ is a simplex and $\sigma'\subset \sigma$, then $\sigma'$ is also a simplex (see \cite{plt}).   If a simplex $\sigma$ 
contains $p+1$ vertices we call it a $p$-simplex.   If every simplex in $K$ is contained in some $D$-simplex we say that $D$ is the dimension of $K$.
Given two abstract simplicial complexes $K$ and $L$, a bijective map $\psi:K^0\rightarrow L^0$ is called a combinatorial isomorphism if it induces a bijection 
of the simplices in $K$ and $L$.

A triangulation is a $3$-dimensional simplicial complex which can be viewed as a collection of tetrahedra together with identifications of 
some pairs of triangles (2-simplices) in the boundaries of the tetrahedra, respecting the colouring, such that any triangle is identified with at most one other triangle.  When we identify triangles we also identify all their subsimplices, i.e.\ their edges (1-simplices) and vertices. 
This point of view will be used and explained in more detail in Section 4 below. Any pair of identified triangles is called an interior triangle of the triangulation while the other triangles are referred to as 
boundary triangles. It should be noted that it is implicit in the notion of a simplicial complex that 
\begin{itemize}
\item[(i)] two triangles contained in the same tetrahedron cannot be identified, 
\item[(ii)] two different triangles in a 
given tetrahedron cannot be identified with two triangles contained in another tetrahedron. 
\end{itemize}
\noindent It is common to speak about these two conditions as regularity conditions and about simplicial complexes as regular triangulations as opposed to singular triangulations, when one or both of these conditions are left out. In this paper we focus on regular triangulations.  Note that the definitions imply that two vertices are 
connected by at most one edge and 3 vertices are contained in at most one triangle.

One may think of a triangulation either as a purely combinatoric object or as a topological space 
embedded in a Euclidean space. In the former case two triangulations are identified if there is a bijective correspondence between their vertices respecting the colouring and the pairwise identifications of triangles. In the latter case two triangulations are identified if there exists a homeomorphism between them mapping simplices to simplices and thus inducing a combinatorial identification. It is a fact, explained in 
e.g. \cite{DJ}, that the two points of view are equivalent.  

We now introduce the basic objects of study in this paper, the two types of causal slices that form the building blocks of the general causal triangulations. 

\noindent 
\begin{definition}\label{def1}
A \emph{causal sphere-slice} $K$ is a triangulation fulfilling the following conditions:

(i) $K$ is homeomorphic to the cylinder $S^2\times [0,1]$

(ii) all monocoloured simplices of $K$ belong to the boundary $\partial K$, such that the red ones belong to one boundary component $\partial K_{\rm red}$ and the blue ones belong to the other component $\partial K_{\rm blue}$.

The set of all causal sphere-slices is denoted by $\cC\cS$.

\end{definition}

\noindent
\begin{definition}\label{def2}
  A {\it causal disc-slice} is a triangulation $K$ fulfilling the following conditions:

(i) $K$ is homeomorphic to the 3-dimensional ball $B^3$

(ii) all monocoloured simplices of $K$ belong to the boundary $\partial K$, such that the red ones form a disc $D_{\rm red}$ and the blue ones form a disc $D_{\rm blue}$, which will be called the boundary discs of $K$. 

The set of all causal disc-slices is denoted by $\cC\cD$.
\end{definition}

We note that the above definitions imply that all the vertices of a causal slice of either type lie on the boundary. Two-coloured edges are sometimes referred to as \emph{timelike edges}. Two-coloured triangles can be of two types, with two red vertices and one blue or vice versa, and are called \emph{forward directed} and \emph{backwards directed} triangles, respectively.

Relaxing the condition in Definition \ref{def2} that $D_{\rm red}$ and $D_{\rm blue}$ are
(homeomorphic to) discs to, say, the requirement that they are deformation retracts of discs 
yields a larger class of causal triangulations, that will not be discussed in detail in this paper (see, however, Section~\ref{sec:6} for some further remarks).   
In the course of the proof of Theorem \ref{thm2} we shall encounter particular examples of such triangulations and we use the same notation as for causal slices without further comment.

Although the main focus of this paper is on causal slices we introduce for the sake of completeness general causal triangulations in the following definition as a layered union of causal slices.   In general these triangulations have interior vertices.

\noindent
\begin{definition}\label{def3} A \emph{causal sphere-triangulation} is a triangulation of the form 
$$
M= \bigcup_{i=1}^{N} K^i\,,
$$ 
where each $K^i$ is a causal sphere-slice such that $K^i$ is disjoint from $K^j$ if $i\neq j$ except that $\partial K_{\rm blue}^i =\partial K_{\rm red}^{i+1}$ for $i=1,\dots,N-1$ as uncoloured 2-dimensional triangulations. 
The boundary components of $M$ are then $\partial K_{\rm red}^1$ and $\partial K_{\rm blue}^N$.

A \emph{causal disc-triangulation} is a triangulation of the form 
$$
M= \bigcup_{i=1}^{N} K^i\,,
$$ 
where each $K^i$ is a causal disc-slice (with boundary discs $D^i_{\rm red}$ and $D^i_{\rm blue}$) such that $K^i$ is disjoint from $K^j$ if $i\neq j$ except that $D^i_{\rm blue} =D_{\rm red}^{i+1}$ for $i=1,\dots,N-1$ as uncoloured 2-dimensional triangulations. The two discs $D^1_{\rm red}$ and $D^N_{\rm blue}$ are called the red and blue boundary disc of $M$, respectively.
\end{definition}

The following lemma states that, apart from the two boundary discs, the boundary of a causal disc-triangulation consists of 
a "timelike cylinder", that is a 2-dimensional causal sphere-triangulation, which is defined as in Definitions \ref{def1} and \ref{def3} with $S^2$ replaced by $S^1$.

\begin{figure}[h]
  \begin{center}
\includegraphics[width=10truecm]{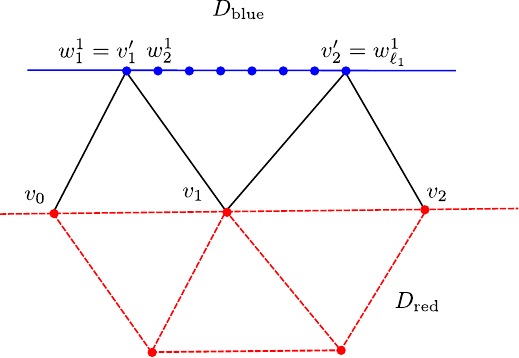}
    \caption{The sequence of triangles on the side of the causal slice $K$ connecting the red and the blue boundary components.}
    \label{Fig1}
      \end{center}
      \end{figure}

\noindent
\begin{lemma}\label{le1}
Let $K$ be a causal disc-slice and denote by $C$ the subcomplex of $\pa K$ consisting of two-coloured triangles. Then $C$ is a 2-dimensional causal sphere-slice with boundary 
$\pa D_{\rm red}\cup \pa D_{\rm blue}$, which will be called the {\it side} of $K$.

More generally, if $M$ is a causal disc-triangulation then the part of the boundary made up of triangles not in the red or blue boundary discs is a 2-dimensional causal sphere-triangulation. 

\end{lemma}
\noindent
{\it Proof.} It is clearly enough to prove the first statement. Choose an orientation of the circle $\pa D_{\rm red}$ and consider a vertex $v_1\in\pa D_{\rm red}$ and its two nearest neighbours $v_0$ and $v_2$ in $\pa D_{\rm red}$ with 
$v_0$ preceding $v_1$ and $v_2$ succeeding $v_1$.  The edges $(v_0v_1)$ and $(v_1v_2)$ are each contained in exactly two triangles of $\pa K$, 
one of which is red while the other one contains a blue vertex $v_1'$, resp.\  $v_2'$, both of which are in $\pa D_{\rm blue}$.
Note that, by the Jordan Curve Theorem, $\pa D_{\rm red}$ divides $\pa K$ into two discs.
It follows that if we denote by $u_1,\ldots ,u_k$ the neighbours of $v_1$ in $D_{\rm red}$ ordered cyclically around $v_1$ such that $u_1=v_0$ 
and $u_k=v_2$, then all other neighbours of $v_1$ in $\pa K$ are blue and constitute a segment $(w_1^1,w_2^1,\ldots , w_{\ell_1}^1)$ of
$\pa D_{\rm blue}$ with $w^1_1=v_1'$ and $w_{\ell_1}^1=v_2'$, see Fig.\ \ref{Fig1}.  Hence, the triangles
$$
(v_0w_1^1v_1), (w_1^1 v_1 w_2^1), \ldots ,(w_{\ell_1 -1}^1 v_1w_{\ell_1}^1), (w_{\ell_1}^1v_1v_2)
$$
make up a segment of a 2-dimensional causal slice.  

Repeating this construction with $v_0,v_1,v_2$ replaced by $v_1,v_2,v_3$, where $v_3$ is the successor 
of $v_2$ in $\pa D_{\rm red}$, we obtain a unique continuation of the segment constructed above by
$$
(w_1^2v_2w_2^2), (w_2^2 v_2 w_3^2), \ldots ,(w_{\ell_2 -1}^2 v_2w_{\ell_2}^2), (w_{\ell_2}^2v_2v_3)
$$
where $w_1^2=w_{\ell_1}^1$ and $(w_1^2, \ldots , w_{\ell_2}^2)$ is a segment of $\pa D_{\rm blue}$.

If the vertices of $\pa D_{\rm red}$ are $v_0,\ldots ,v_n$ 
then after $n$ steps we obtain a segment of a causal slice with horizontal edges in
$\pa D_{\rm red}\cup \pa D_{\rm blue}$ and whose first and last triangle share the vertex $v_0$.   Finally this segment can be completed to a 2-dimensional causal slice $C$ 
by adjoining the triangles which contain $v_0$ and have one blue edge 
(in $\pa D_{\rm blue}$) and two non-coloured edges.  By construction $C$ has boundary
$\pa D_{\rm red}\cup \pa D_{\rm blue}$ and evidently constitutes all of $\pa K\setminus {\rm int} (D_{\rm red}\cup D_{\rm blue})$.
\hfill $\Box $

\medskip

The next two propositions are elementary and ensure the existence of causal disc-slices and sphere-slices with prescribed boundary discs, respectively boundary components.

\begin{figure}[h]
  \begin{center}
\includegraphics[width=6truecm]{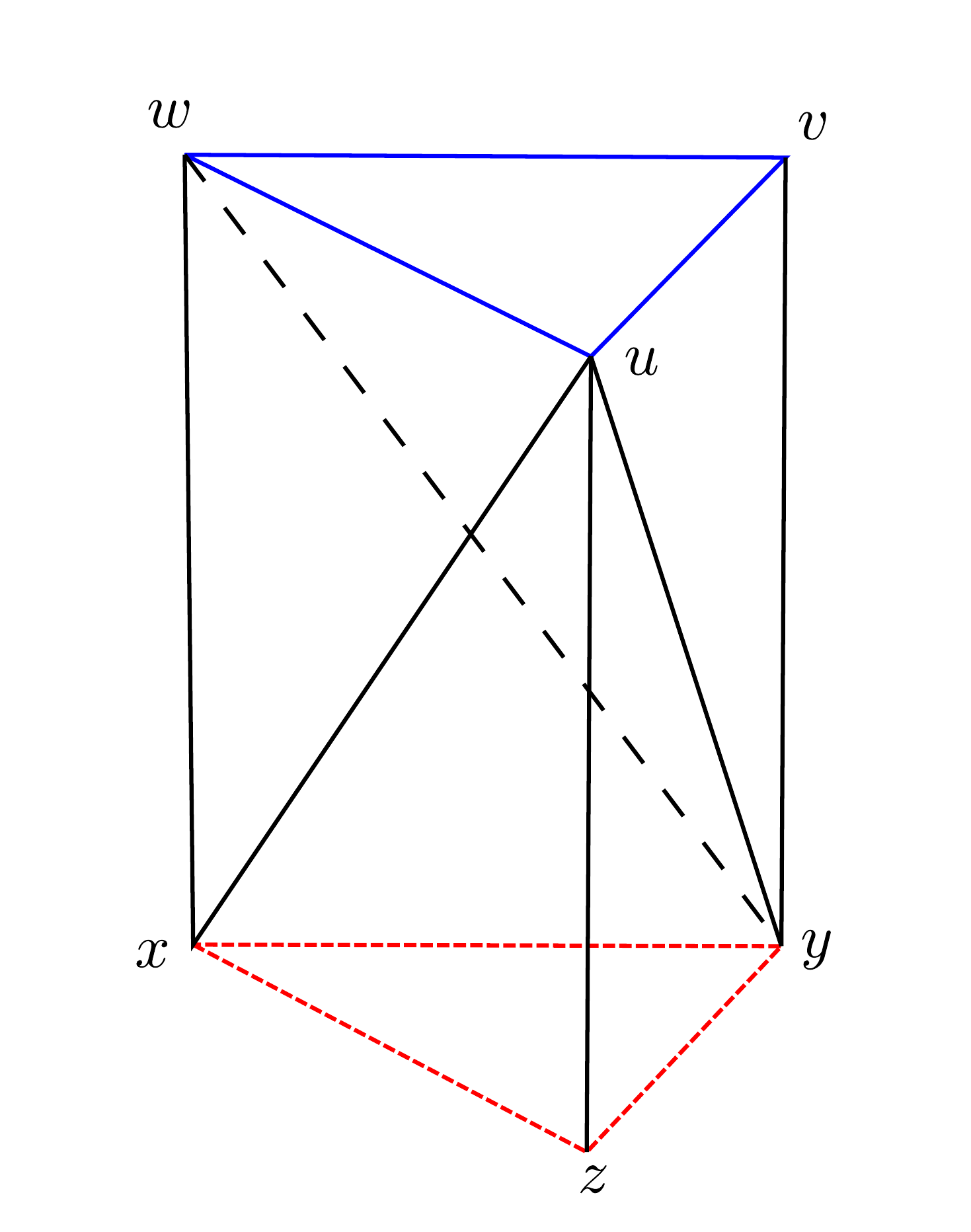}
    \caption{A prism made up of 3 tetrahedra $(xyzu)$, 
    $(uvwy)$, and $(xyuw)$.}
    \label{Fig2}
      \end{center}
      \end{figure}

\noindent 
\begin{proposition}\label{prop1}
Given two triangulated discs $D_1$ and $D_2$ there exists a causal disc-slice $K$ such that 
$D_{\rm red}=D_1$ and $D_{\rm blue}=D_2$.

\end{proposition}
\noindent
{\it Proof.}  We give an inductive argument. Suppose first that $D_1$ and $D_2$ are triangles $\Delta_{\rm red}$ and $\Delta_{\rm blue}$, respectively. In this case, we can choose $K$ to be the prism, depicted in Fig.\ \ref{Fig2}, made up of one $(1,3)$ tetrahedron, 
one $(3,1)$ tetrahedron and one $(2,2)$ tetrahedron.

Now assume $K$ exists for some given $D_1$ and $D_2$. If $e$ is an arbitrary edge in 
$\pa D_1=\pa D_{\rm red}\subset \pa K$ one can glue a 
tetrahedron of type $(3,1)$ to $K$ along the unique triangle in the side of $K$ which contains $e$.  Thus we obtain a causal 
disc-slice $K'$ whose red boundary disc has an extra triangle compared to that of $K$ but the blue boundary disc is the same.

\begin{figure}[h]
  \begin{center}
\includegraphics[width=11truecm]{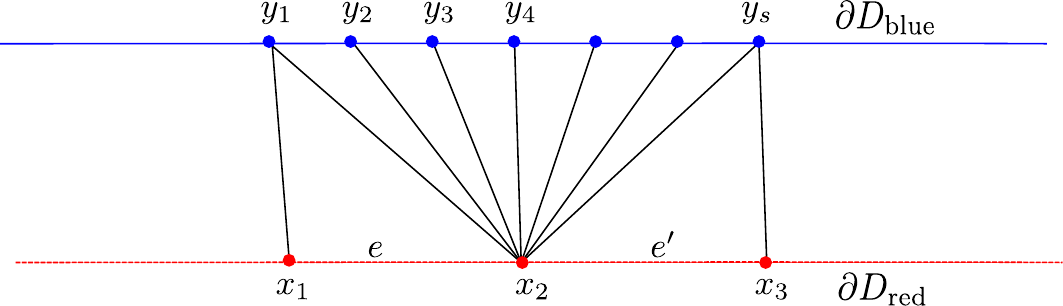}
    \caption{Triangles in the side of a causal disc slice.}
    \label{Fig3}
      \end{center}
      \end{figure}

Similarly, given two neighbouring edges $e=(x_1x_2)$ and $e'=(x_2x_3)$ in 
$\pa D_1$, see Fig.\ \ref{Fig3}, one can first identify $e$ 
and $e'$ and then glue a sequence of $(2,2)$ tetrahedra sharing the identified $e$ and $e'$ and whose blue edges are 
$$
(y_1y_2), (y_2y_3), \ldots , (y_{s-1}y_s)
$$ 
with the notation of Fig.\ \ref{Fig3}. The resulting triangulation 
$K''$ is a causal disc-slice with red boundary disc $D'_{\rm red}$ obtained from $D_{\rm red}$ by identifying $e$ and $e'$
while $D_{\rm blue}$ is unchanged.

Similar constructions can of course be made with $D_{\rm blue}$ replacing $D_{\rm red}$.
Starting with an arbitrary triangle, it is well known (see e.g.\ \cite{dur} or \cite{bz}) that $D_1$ (and similarly $D_2$) can be constructed by repeated application of the process of either gluing on a triangle or by identifying two neighbouring boundary edges as described. 
Hence, the existence of $K$ follows from the preceding discussion by induction.
 Note that since $D_1$ and $D_2$ are regular triangulations so is the causal disc-slice $K$.
\hfill$\Box$

\noindent 
\begin{proposition}\label{prop2}
 Given two triangulated 2-spheres, $S_1$ and $S_2$, there exists a causal 
sphere-slice $K$ such that $S_1$ is
the red boundary of $K$ and the blue boundary is $S_2$.

\end{proposition}

\begin{figure}[h]
  \begin{center}
 \includegraphics[width=10truecm]{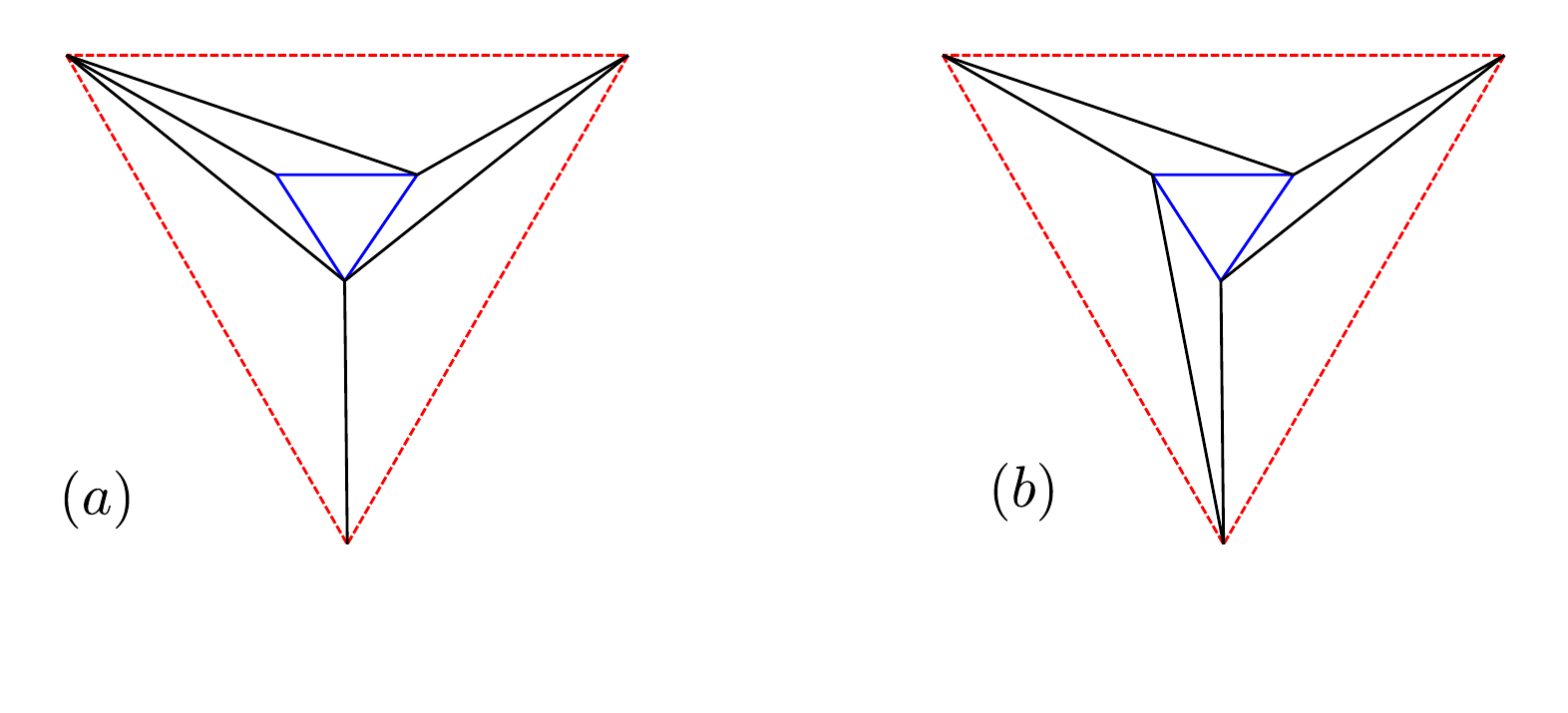}
    \caption{The two possible triangulations of the side of a disc-slice whose red and blue boundary components are both a single triangle.}
    \label{Fig4}
      \end{center}
      \end{figure}

\noindent
{\it Proof.}  We remove one triangle from $S_1$ and another one from $S_2$.  Then we obtain two triangulated discs $D_1$ and 
$D_2$.  By Proposition \ref{prop1} there exists a causal disc-slice $K'$ with boundary discs $D_{\rm red}=D_1$ and $D_{\rm blue}=D_2$. 
The side of $K'$, $C$, is a causal 2-dimensional disc-slice whose boundary components are triangles.  There are two different
possibilities for $C$, up to combinatorial equivalence, see Fig.\ \ref{Fig4}.

Note that we may assume that no pair of neighbouring triangles in $C$ belongs to the same tetrahedron.  Indeed, if 
the two triangles are both forward (or both backward) directed, this follows from the fact that 
otherwise the removed triangle in either $S_1$ or $S_2$ would be glued to another triangle along two edges contradicting the regularity of $S_1$ and $S_2$. 
If one backward triangle and one forward triangle in $C$ belong to the same tetrahedron in $K'$ it must be a 
(2,2)-tetrahedron which can be removed from $K'$ without changing the red and blue boundary discs (only an edge in $C$ 
gets flipped).  Since there are only finitely many (2,2)-tetrahedra in $K'$ the claim follows. 

If now $C$ is of type (a) in Fig.\ \ref{Fig4}, then we can glue to $K'$ a prism of the form indicated 
in Fig.\ \ref{Fig2} to obtain $K$ as desired.
If $C$ is of type (b) we can glue onto the prism of Fig.\ \ref{Fig2} a 
(2,2)-tetrahedron 
to obtain 
a regular prism whose side $C'$ is of the same type as $C$ 
and hence can be glued onto $K'$ to obtain $K$ as desired.  
\hfill $\Box$

\section{Disc-slices and midsections}
Given a causal disc-slice $K$ we define its midsection $S_K$ as in \cite{DJ} for a causal sphere-slice. More explicitly, we view $K$ as being embedded in a Euclidean space and consider a tetrahedron 
$t=(v_1v_2v_3v_4)$ in $K$ with vertices $v_1, v_2, v_3, v_4$, i.e. $t$ consists of the points of the form 
\begin{equation}\label{height}
x= s_1 v_1 + s_2 v_2 +s_3 v_3 + s_4 v_4\,, 
\end{equation}
where $s_1, s_2, s_3, s_4 \geq 0$ and $s_1 + s_2 + s_3 + s_4 =1$. Letting the height $h(x)$ be defined as the sum of the coefficients 
of the red vertices in \eqref{height} the 2-cell $F$ corresponding to $t$ is defined as the set of points 
$x$ in $t$ of height $h(x)=\oh$. If $t$ is of type (3,1) then $F$ is a triangle whose edges by convention are coloured red; if $t$ is of type (1,3) then $F$ is likewise a triangle whose edges are coloured blue; if $t$ is of type (2,2) then $F$ is a quadrangle 
each of whose boundary edges are contained in exactly one boundary triangle of $t$ containing a monocoloured edge of $t$; 
by convention each edge of $F$ inherits the colour of the corresponding monocoloured edge in $t$. 
It is easy to show (see \cite{DJ} for details) that the 2-cells so obtained define a 2-dimensional cell complex homeomorphic to $D^2$ with edges coloured red or blue in such a way that triangles are monocoloured while quadrangles are two-coloured with opposite edges of the same colour. Moreover, this cell complex is uniquely defined up to combinatorial isomorphism and is called the \emph{midsection} of $K$, denoted by 
$S_K$. 

By construction each vertex $a$ of $S_K$ is contained in a unique edge $e_a$ of $K$ whose endpoints have different colours and vice versa. Similarly, any red (blue) edge of $S_K$ is contained in a unique two-coloured triangle of $K$ which contains a red (blue) edge in $\partial K$, and vice versa. Finally, each 2-cell of $S_K$ is contained in a unique tetrahedron of $K$, the tetrahedron being of type (3,1), (2,2) or 
(1,3) depending on whether the cell is a red triangle, a quadrangle, or a blue triangle, respectively.

In a 2-dimensional coloured cell complex as described we shall use the notation $\langle ab\rangle$ and $\langle abc\rangle$ for edges and triangles with vertices $a,b$ and $a,b,c$, respectively, whereas a 2-cell with cyclically ordered vertices $a,b,c,d$, such that $\langle ab\rangle$ and $\langle cd\rangle$ are red edges, will be denoted by $\langle abcd\rangle$. Note that with this convention we have, e.g., $\langle abcd\rangle$ =$\langle dcba\rangle$ =$\langle cdab\rangle$. 

By a red, resp. blue, path in $S_K$ we mean a sequence $e_1,\dots,e_k$ of red, resp. blue, edges such that 
$e_i=\langle a_ia_{i+1}\rangle$ for each $i=1,\dots, k-1$, and some vertices $a_1,\dots,a_k$. In this case we say that the path connects $a_1$ and $a_k$. The path is called simple if either the vertices 
$a_1,\dots,a_{k}$ are all different or if $a_1,\dots,a_{k-1}$ are different while $a_1=a_k$, in which case the path is said to be closed.

\noindent  
\begin{remark} \label{rem1} 
{\rm With the notation just introduced it follows from the definition of $S_K$ 
that if $a$ and $b$ are two vertices of $S_K$ then the red (blue) endpoints of $e_a$ and $e_b$ are identical if $a$ and $b$ 
are connected by a blue (red) path. Indeed, if $e=\langle ab\rangle$ is a red edge of $S_K$, then 
$e$ is contained in a 
triangle in $K$ two of whose edges are $e_a$ and $e_b$ sharing a blue vertex.
Evidently, the claim follows from this, and similarly if $e=\langle ab\rangle$ is blue.

 The converse statement that 
$a$ and $b$ are connected by a red (blue) path if $e_a$ and $e_b$ share a blue (red) endpoint also holds as a consequence of the 
proof of Theorem \ref{thm1} below. }

\end{remark}

Let us now note that since the monocoloured 
boundary discs of $K$ are non-empty, it follows that $S_K$ contains at least one triangle of each colour. For the same reason the boundary of $S_K$ must contain edges of both colours. As a consequence, the boundary 
of $S_K$ consists of a (cyclically ordered) alternating sequence of monocoloured paths which we shall call \emph{boundary arcs}. 
The following few lemmas describe 
properties satisfied by any midsection $S_K$.

\begin{lemma}
\label{le2}
 Two different vertices in a midsection $S_K$ cannot be connected by both a red and a blue path.

\end{lemma}

\noindent {\it Proof.} If two vertices $a$ and $b$ are connected by both a red and a blue path, then the two edges $e_a$ and $e_b$ in $K$ have identical endpoints and hence are identical, which implies $a=b$. 
\hfill $\Box$

\medskip

Since $\pa D_{\rm red}$ is homeomorphic to $S^1$, there does not exist a sequence  of quadrangles $q_1,q_2,\ldots ,q_k$ in 
$S_K$ such that $q_i$ and $q_{i+1}$ share a red edge for $i=1,2,\ldots ,k-1$ and the red edges in $q_1$ and $q_k$, not shared with $q_2$ or $q_{k-1}$, belong to $\pa D_{\rm red}$.   A sequence of quadrangles as we have described will be called a 
{\it red path of quadrangles} connecting edges in $\pa D_{\rm red}$.  The corresponding statement with red replaced by blue 
is of course also true.   We define a closed path of quadrangles (red or blue) analogously.  Obviously there cannot exist a closed path 
of quadrangles in $S_K$ because then $K$ would contain a blue or a red edge in its interior.
The absence of paths of quadrangles just described follows from the following more restrictive conditions.

\begin{lemma}\label{le3}
Let $S_K$ be a midsection, let $\rho$ denote a closed simple red (blue) 
path in $S_K$ and let $\mu$ be a simple red (blue) path connecting two vertices belonging to different 
blue (red) arcs in the boundary $\partial S_K$. Then the following hold:

i) The interior of $\rho$, i.e. the component of $S_K\setminus\rho$ not containing any boundary edges, contains solely red (blue) edges.

ii) The two endpoints of $\mu$ are the two endpoints of a red (blue) boundary arc.  

\end{lemma}
\noindent {\it Proof.} i) It is sufficient to consider the case when $\rho$ is red. If $\rho$ encloses a blue edge it clearly also encloses a blue triangle. Since the boundary of $S_K$ contains blue edges the exterior of $\rho$ likewise contains a blue triangle. Evidently, all red paths connecting a vertex in an interior blue triangle with a vertex in an exterior blue triangle must intersect $\rho$. For $K$ this 
means that $D_{\rm blue}$ consists of two nontrivial subcomplexes sharing a single vertex (namely the common blue endpoint of the edges $e_a$ for $a\in\rho$) which contradicts the fact that $D_{\rm blue}$ is a disc.

ii)  Assume $\mu$ is red and connects two vertices $a$ and $b$ belonging to two different blue boundary arcs $\alpha_1$ and $\alpha_2$. If the conclusion of ii) does not hold one sees that the common blue endpoint of the edges $e_a$ and $e_b$ in $K$ would separate $\partial D_{\rm blue}$ into two non-trivial parts sharing only this vertex. This contradicts the fact that $\partial D_{\rm blue}$ is a simple closed curve. Obviously, a similar argument applies when $\mu$ is blue. 
\hfill $\Box$

\begin{lemma}\label{le4}
Let $e=\langle ab\rangle$ and $f=\langle a'b'\rangle$ be two disjoint blue (red) edges in the midsection $S_K$.  Suppose $a$ and $a'$ as well as $b$ and $b'$ are connected by a red (blue) path.  Then there exists a blue (red) path of quadrangles connecting $e$ and $f$.

\end{lemma}
\noindent
{\it Proof.} It suffices to prove the Lemma for blue edges.  
Let $\Delta_e$ and $\Delta_f$ be the two-coloured triangles of $K$ that contain the blue edges $e$ and $f$, respectively.  
Then $\Delta_e$ and $\Delta_f$ share a blue edge
$(xy)$ in the blue boundary of $K$ and also have red vertices $v_e$ and $v_f$ in $\pa K$.  
Since $e$ and $f$ are disjoint $v_e\neq v_f$.  If $(xy)$ belongs to $\pa D_{\rm blue}$, the star of $(xy)$ consists of a sequence
$(xyv_1v_2), (xyv_2v_3),\ldots ,(xyv_{k-1}v_k)$ of (2,2)-tetrahedra sharing $(xy)$ and an additional 
(1,3)-tetrahedron $(xyzv_k)$.  We have $v_e,v_f\in\{ v_1,\ldots ,v_k\}$ and the claim follows since each (2,2)-tetrahedron corresponds to a quadrangle in the midsection.  If $(xy)$ is an interior edge in $D_{\rm blue}$ then the first (2,2)-tetrahedron above has to be preceded by a (1,3)-tetrahedron $(xyz'v_1)$ and then the rest of the argument is unchanged.
\hfill $\Box$

\medskip
\noindent
\begin{proposition}\label{prop3}
The midsection $S_K$ of a causal disc-slice $K$ determines $K$ uniquely up to 
combinatorial equivalence.

\end{proposition}

\noindent
{\it Proof.} This follows by arguments identical to those in \cite{DJ} for causal sphere-slices. 
\hfill $\Box$

\noindent
\begin{definition}
We let $\cS\cD$ denote the set of all coloured cell complexes $S$ homeomorphic to a disc (with cells as described previously) which have at least one triangle of each colour and satisfy the following conditions:
\begin{itemize}

\item[($\alpha$)] No pair of different vertices in $S$ are connected by both a red and a blue path. 

\item[($\beta_1$)] Each closed simple red (blue) path encloses solely red (blue) triangles in its interior.

\item[($\beta_2$)] Considering the division of $\pa S$ into red and blue arcs, there is no red (blue) path connecting
two vertices belonging to different blue (red) arcs unless they are the two endpoints of a red (blue) arc.

\item[($\gamma$)]  If $e=\langle ab\rangle$ and $f=\langle cd\rangle$ are two disjoint blue (red) edges in $S$ such that $a$ and $c$ as well as $b$ and $d$ are connected by a red (blue) path, then there exists a blue (red) path of quadrangles connecting $e$ and $f$.

\end{itemize}
\end{definition}

\begin{remark}\label{rem2} {\rm
As mentioned previously, condition $(\beta_1)$ implies the absence of closed red (blue) paths of quadrangles. Indeed, the outer blue (red) boundary component of such a path would violate $(\beta_1)$.  Similarly, condition $(\beta_2)$ implies that no two different edges in $\pa S$ can be connected by a red or blue path of quadrangles. Likewise, it follows that two different edges in the same triangle cannot be connected by a path of quadrangles, since it would contradict ($\alpha$).
One can demonstrate by explicit examples that property ($\gamma$) does not follow from the first three properties. 

It was noted in \cite{al2,al3} that the dual graphs that arise in the matrix model formulation of 3-dimensional causal triangulations and 
correspond to 3-dimensional simplicial manifolds satisfy some extra conditions that are closely related to $(\alpha), (\beta_1)$ and $(\gamma)$ above.}

\end{remark}

\begin{definition}\label{delta} 
We say that a 2-dimensional coloured cell complex satisfies condition $(\delta)$ if for any pair of distinct 
red, resp.\ blue, triangles it is not possible to join their vertices pairwise by blue, resp.\ red, paths.
\end{definition}

%
 
\noindent
\begin{lemma}\label{le6}
Condition ($\delta$) holds for all $S\in\cS\cD$.

\end{lemma}

\noindent
{\it Proof.} 
Given the two triangles, let us assume they are red and that three blue paths exist connecting $a$ to $a'$,
 $b$ to $b'$, and $c$ to $c'$. It follows easily from property ($\alpha$) that the two triangles must be disjoint. Using ($\gamma$) there exists a (non-trivial) path $\rho_{ab}$ of quadrangles connecting $\br ab\kt$ to $\br a'b'\kt$. Similarly, a path $\rho_{ac}$ of quadrangles exists connecting 
$\br ac\kt$ to $\br a'c'\kt$. Each of these two paths contains a blue path connecting $a$ to $a'$ which do not intersect each other (although they may touch at some vertices or edges). Hence they define a closed curve, whose interior consists of blue triangles by ($\beta_1$). In particular, all edges of those blue triangles as well as all red edges of $\rho_{ab}$ and $\rho_{ac}$ are interior edges of $S$, and the exterior of the curve contains the two original triangles and also the path $\rho_{bc}$ of quadrangles connecting $\br bc\kt$ to $\br b'c'\kt$, whose existence again follows from ($\gamma$). Similarly, considering the closed blue curves determined by $\rho_{ab}$ and $\rho_{bc}$, respectively $\rho_{bc}$ and $\rho_{ac}$, we conclude that all edges are interior in $S$ which contradicts the fact that $S$ is a disc.
\hfill $\Box$

\medskip

Lemmas \ref{le2} - \ref{le6} together with Proposition \ref{prop3} show that the mapping
$\psi : K \mapsto S_K$ is a well-defined injective map from the set $\cC\cD$ of causal disc-slices 
into $\cS\cD$.  We now aim to prove the following main result of the present paper.

\noindent
\begin{theorem}\label{thm1}
The mapping $\psi :K\mapsto S_K$ is bijective from the set $\cC\cD$ 
of causal disc-slices onto the set $\cS\D$ of coloured 2-dimensional cell complexes.

\end{theorem}

\section{Proof of the main result} 

In this section we prove Theorem \ref{thm1}.
The strategy is to show first that from any $S\in \cS\cD$ we can construct a unique simplicial complex.
We then proceed to show that this simplicial complex has the topology of a ball and is actually a causal disc-slice.
The midsection of this causal slice is the coloured cell complex we started with.

Let $S\in\cS\cD$ be given.  In order to construct the corresponding $K\equiv K_S\in\cC\cD$ we start by 
associating to each vertex $a\in S$ a pair of new vertices $r_a, b_a$ which will form the vertex set $K_S^0$ 
of $K_S$ with the following identifications:
\begin{itemize}
\item[($\star$)] 
$r_a=r_b$, resp. $b_a=b_b$, if $a$ and $b$ are connected by a blue, resp. red, path,
\end{itemize}
\noindent
where $a,b$ are arbitrary vertices in $S$.
Thus $K_S^0$ consists of all the vertices $r_a,b_a$ with $a\in S$ subject to the identifications ($\star$). By definition we attach the colour red to the vertex $r_a$ while $b_a$ is coloured blue.  

The set $K_S^3$ of coloured tetrahedra is obtained from the collection of 
2-cells of $S$ as follows:  
for each red triangle $\Delta =\langle abc\rangle$ let $t_\Delta$ be the (3,1)-tetrahedron $(r_ar_br_cb_a)$
where we notice that all 4 vertices are different by ($\alpha$) and $b_a=b_b=b_c$;
similarly, if $\Delta$ is a blue triangle, let $t_\Delta =(b_ab_bb_cr_a)$; finally, for the quadrangle $q=\br abcd\kt$, we let $t_q$ be the (2,2)-tetrahedron $(r_ar_bb_ab_c)$;  again it follows from 
($\alpha$) that the four vertices are different and $t_q$ depends only on the quadrangle $q$.

Thus to each 2-cell $F$ of $S$ there corresponds a tetrahedron $t_F$ with vertices in $K_S^0$.  This defines an abstract 
3-dimensional coloured simplicial complex $K_S$ whose edges and triangles are obtained as sub-simplices of the tetrahedra.

Let us first verify that $F\mapsto t_F$ is bijective between 2-cells in $S$ and tetrahedra in $K_S$.  By definition the mapping is
surjective.  Consider two 2-cells $F$ and $F'$ such that $t_F=t_{F'}$.  Clearly $F$ and $F'$ are both triangles with the same colour or they are both quadrangles.  

Suppose $F=\br abc\kt$ and $F'=\br a'b'c'\kt$ are, say, red triangles.  Then $t_F=t_{F'}$ means that 
$b_a=b_b=b_c=b_{a'}=b_{b'}=b_{c'}$ and $\{r_a,r_b,r_c\}=\{r_{a'},r_{b'},r_{c'}\}$.  By ($\alpha$) this implies that $\{a,b,c\}=\{a',b',c'\}$ and hence $\br abc\kt =\br a'b'c'\kt$.

If $F=\br abcd\kt$ and $F'=\br a'b'c'd'\kt$ are quadrangles then 
$t_F=t_{F'}$ implies that $\{r_a,r_b\} = \{r_{a'}, r_{b'}\}$ and $\{b_a, b_c\}=\{b_{a'},b_{c'}\}$. 
 Using ($\alpha$) it is then straightforward to check that 
$\br abcd\kt$ is equal to $\br a'b'c'd'\kt$.  We have thus established that the tetrahedra of $K_S$ are labelled by the 2-cells of $S$.

Next let us consider the triangles in $K_S$.  These fall into four disjoint classes:
\begin{itemize}

\item[(i)] Red triangles $(r_a r_b r_c)$ where $\br abc\kt$ is a red triangle in $S$.
\item[(ii)] Blue triangles $(b_a b_b b_c)$ where $\br abc\kt$ is a blue triangle in $S$.
\item[(iii)] Triangles $(r_a r_b b_a)$ where $\br ab\kt$ is a red edge in $S$.
\item[(iv)]  Triangles $(b_a b_b r_a)$ where $\br ab\kt$ is a blue edge in $S$. 

\end{itemize}
Using property ($\delta$), which holds by Lemma \ref{le6}, we 
see that red triangles in $K_S$ corresponding to different red triangles in $S$ are different.
Such triangles are not shared by different tetrahedra and therefore lie in the boundary of $K_S$.  The corresponding 
statement about blue triangles is also clearly true.

Next consider a triangle $(r_a r_b b_a)$ where $\br ab\kt$ is a red edge in $S$. By definition of $t_F$ it holds that, if $\br ab\kt$ belongs to a 2-cell $F$ in $S$, then $(r_ar_b b_a)$ belongs to the boundary of $t_F$.  

Conversely, suppose the triangle $(r_a r_b b_a)$ belongs to the boundary of a tetrahedron $t_F$.
If $F$ is a red triangle $\br a'b'c'\kt$, then $(r_a r_b b_a)$ 
equals one of the triangles $(r_{a'} r_{b'} b_{a'})$, $(r_{a'} r_{c'} b_{a'})$, $(r_{b'} r_{c'} b_{a'})$.
By ($\alpha$) this implies that $\{a,b\}$ equals one of $\{ a',b'\}$,
 $\{ a',c'\}$, $\{ b',c'\}$ and hence the edge $\br ab\kt$ equals one of the edges $\br a'b'\kt$,
 $\br a'c'\kt$, $\br b'c'\kt$.  
Thus, $\br ab\kt$ belongs to $F$.
Similarly, if $F$ is a quadrangle $\br a'b'c'd'\kt$, it follows that
$(r_a r_b b_a)$ equals $(r_{a'}r_{b'}b_{a'})$ or   $(r_{a'}r_{b'}b_{c'})$. 
By ($\alpha$) this implies that $\br ab\kt=\br a'b'\kt$ or $\br ab\kt=\br c'd'\kt$ and hence $\br ab\kt$ is 
an edge in $F$.

From these observations follows that two tetrahedra $t_F$ and $t_{F'}$ share a triangle 
$\Delta =(r_ar_bb_a)$ (or $\Delta =(b_ab_br_a)$) if and only if $F$ and $F'$ share the 
edge $\br ab\kt$.  In particular, the interior triangles in $K_S$ 
are labelled by the interior edges in $S$.  The boundary triangles in $K_S$ are labelled by the boundary edges of $S$ together with the monocoloured triangles in $S$ which label the monocoloured boundary triangles in $K_S$.

We next consider the edges in $K_S$. In particular, we want to show that monocoloured edges lie in $\pa K_S$.  Let
$(r_ar_b)$ be a red edge in $K_S$.  This means that $a,b$ can be assumed to belong to some 2-cell $F$ in $S$ such that $\br ab\kt$ is  a red edge in $F$.  If $F=\br abc\kt$ is a red triangle it follows from the preceding paragraph that $(r_ar_b)$ belongs to $\pa K_S$ since it belongs to the red triangle $(r_ar_br_c)$.
Alternatively, if $\br ab\kt$ belongs to a quadrangle $q$, then by ($\beta_1$) and ($\beta_2$)
we have that $q$ belongs to a red path of quadrangles connecting either two red triangles
or a red triangle and a red boundary edge in $S$.  This shows that $(r_ar_b)$ is an edge in a red triangle in $\pa K_S$.
Applying similar arguments to blue edges in $K_S$ shows that all monocoloured edges lie in the boundary.

The correspondence $a\to (r_ab_a)$ between vertices in $S$ and the two-coloured edges of $K_S$ is bijective by property ($\alpha$). Moreover, $(r_ab_a)$ belongs to $\pa K_S$ if and only if $a$ belongs to $\pa S$, since it belongs to the two boundary triangles corresponding to the edges in $\pa S$ incident on $a$.   

Finally, consider a red vertex $x$ in $K_S$.  It belongs to some tetrahedron $t_F$ where $F$ is a 2-cell in $S$.  If $F$ is a quadrangle or a red triangle, then $x$ belongs to a red edge and hence, by the preceding paragraph, it belongs to $\pa K_S$.  If $F$ is a blue triangle 
$\br abc\kt$ then $x=r_a=r_b=r_c$.  
Since we assume that there is a least one triangle of each colour in $S$ we can pick
a path in $S$ starting at $a$ and ending at vertex $a'$ in a red triangle $\br a'b'c'\kt$.  There is a first vertex $d$ in the path which is contained in a red edge and we have $x=r_d$.  Hence, $x$ belongs to $\pa K_S$.  An identical argument shows that a blue vertex in $K_S$ 
is necessarily contained in the boundary.   This completes the argument that all 
monocoloured simplices in $K_S$ belong to the boundary. 

The next thing to consider is the structure of $\pa K_S$.  We have seen above that there is a one-to-one correspondence between the boundary edges in $S$ and the two-coloured triangles in $\pa K_S$.  It follows that these triangles form a sequence
$\Delta_1,\ldots ,\Delta_k$ corresponding to the boundary edges $e_1,\ldots ,e_k$ in $\pa S$ ordered cyclically.
Setting $e_i=\br a_ia_{i+1}\kt$, where $a_1, \ldots ,a_k$ are the cyclically ordered vertices of 
$\pa S$ (with $a_{k+1}=a_1$), we see that
$\Delta_i$ and $\Delta_{i+1}$ share the edge $(r_{a_{i+1}}b_{a_{i+1}})$. 
By ($\beta_2$) the triangles
$\Delta_1,\ldots ,\Delta_k$ make up a 2-dimensional causal slice $C$, homeomorphic to the cylinder $S^1\times [0,1]$, whose red and blue 
boundary circles are denoted $\pa C_{\rm red}$ and $\pa C_{\rm blue}$.  
At this stage we need the following lemma whose proof we will postpone a little.

\noindent
\begin{lemma}\label{le7}
The simplicial complex $K_S$ is homeomorphic to the 3-ball so its boundary is homeomorphic to 
the 2-sphere.
\end{lemma}

By the above lemma $\pa K_S\setminus C$ consists of two discs $D_{\rm red}$ and $D_{\rm blue}$ whose boundaries are
$\pa C_{\rm red}$ and $\pa C_{\rm blue}$.  We claim that $D_{\rm red}$ is made up exactly of the red triangles in $\pa K_S$ and similarly for $D_{\rm blue}$.  Indeed, since $D_{\rm red}$ is a 2-dimensional pseudomanifold any triangle in $D_{\rm red}$ can be edge-connected to $\pa D_{\rm red}=\pa C_{\rm red}$ by a sequence of triangles in $D_{\rm red}$.  The triangles in $D_{\rm red}$ 
are monocoloured and are therefore either red or blue.  But a 
blue triangle cannot be glued to a red triangle so $D_{\rm red}$
consists of red triangles.  These must be all the red triangles in $\pa K_S$ since $D_{\rm blue}$ consists solely of blue triangles.  
We have therefore shown that $K_S$ is a causal disc slice.

Now it is not hard to verify that the midsection of $K_S$ is combinatorially isomorphic 
to the original midsection $S$ used to construct $K_S$. We already noted above that the vertices of $S$ are in bijective correspondence with the two-coloured edges in $K_S$. By the definition of the midsection $S'$ of $K_S$ its vertices are the midpoints of the two-coloured edges of $K_S$, and hence can be labelled by the vertices of $S$. In this way the midsection of $t_F$ clearly gets identified with $F$ and the edge in $S'$ corresponding to an interior triangle $(r_ar_bb_a)$, respectively $(b_ab_br_a)$, of $K_S$ is the red, respectively blue, edge $\br ab\kt$ in $S$. Hence, the correspondence between vertices induces bijective correspondences between 2-cells and 1-cells as well, and so $S$ and $S'$ are combinatorially isomorphic. This completes the proof of Theorem \ref{thm1}.      \hfill $\Box$
 
 \medskip

It remains to prove Lemma \ref{le7}. For this purpose an alternative 
construction of $K_S$ by a gluing procedure is a useful tool.  We begin by explaining this construction.

For each 2-cell $F$ in $S$ and each vertex $a$ of $F$ we define a red vertex $r_a^F$ and a blue 
vertex $b_a^F$ such that $r_a^F=r_b^F$, resp.\ $b_a^F=b_b^F$, if $\br ab\kt$ is a blue, resp.\ red, edge in F. In this way four different vertices are defined for each 2-cell $F$ and they in turn define a coloured (abstract) tetrahedron $\tau_F$ considered as a simplicial complex including its subsimplices. As before $\tau_F$ is of type $(3,1),(2,2)$, or $(1,3)$ depending on whether $F$ is a red triangle, a quadrangle, or a blue triangle, respectively. Without further identifications the tetrahedra so defined are pairwise disjoint and we note that the definition of $K_S$ can be reformulated by stating that $K_S$ is obtained from the collection of tetrahedra $\tau_F$ labeled by the 2-cells of $S$ by imposing the identifications of simplices implied by the relations 

\begin{itemize}
\item[($\star\star$)] 
$r_a^{F_1}=r_b^{F_2}$, resp. $b_a^{F_1}=b_b^{F_2}$, if either $a=b$ or $a$ is connected to 
$b$ by a blue, resp.\ red, path,
\end{itemize}
\noindent
where $a,b$ are arbitrary vertices in $S$ and $F_1, F_2$ are arbitrary 2-cells in $S$ containing $a$ and $b$, respectively.

 We next show that $K_S$ can equivalently be obtained by applying a suitable gluing procedure to the collection of 
 tetrahedra defined above.
Given two 2-cells $F_1$ and $F_2$ sharing a red edge $\br ab\kt$ we have that $\tau_{F_1}$ contains the triangle $(r_a^{F_1}r_b^{F_1}b_a^{F_1})$ and $\tau_{F_2}$ contains the triangle 
$(r_a^{F_2}r_b^{F_2}b_a^{F_2})$. We say that $\tau_{F_1}$ is glued to $\tau_{F_2}$ along $\br ab\kt$ if 
$r_a^{F_1}$ is identified with $r_a^{F_2}$, $r_b^{F_1}$ is identified with $r_b^{F_2}$, and $b_a^{F_1}$ is identified with $b_a^{F_2}$ in $\tau_{F_1}$ and $\tau_{F_2}$  and their subsimplices. Similarly, gluing along a blue interior edge in $S$ is defined. In this way, given an interior edge $\br ab\kt$ 
of $S$, the tetrahedra $\tau_{F_1}$ and $\tau_{F_2}$ corresponding to the 2-cells $F_1$ and $F_2$ 
sharing $\br ab\kt$ can be glued along $\br ab\kt$.  

We now define the simplicial complex $K'_S$ by imposing the identifications of the simplices in the collection $\{\tau_F\}$ implied by gluing pairs of tetrahedra as described along all interior 
edges of $S$. We claim that $K_S=K'_S$. In order to prove this we need to verify that the identifications of vertices implied by the gluing conditions are identical to those given 
by $(\star\star)$.

First, consider $r_a^{F'}$ and $r_a^{F''}$ where $a$ is a vertex in two different 2-cells $F'$ and $F''$. Since $S$ is a manifold 
there exist 2-cells $F_1, F_2,\dots,F_n$ such that $F_1=F'$ and $F_n=F''$ and $F_i$ and $F_{i+1}$ share an edge $e_i$ 
containing $a$ for each $i=1,\dots,n$. From the gluing of $\tau_{F_i}$ and $\tau_{F_{i+1}}$ along $e_i$  
it follows that $r_a^{F_i}= r_a^{F_{i+1}}$ and $b_a^{F_i}= b_a^{F_{i+1}}$ for $i=1,\dots,n-1$, and hence we conclude that $r_a^{F'}= r_a^{F''}$ and $b_a^{F'}= b_a^{F''}$.   

Next consider two different vertices $a$ and $b$ in $S$ and assume they are connected by a blue path with edges $\br a_1a_2\kt, \br a_2a_3\kt,\dots, \br a_ma_{m+1}\kt$, where $a_1=a$ and $a_{m+1}=b$. Choosing arbitrary 2-cells $F_1,\dots, F_m$ such that $\br a_ia_{i+1}\kt$ belongs to $F_i, i=1,\dots,m$, we have that $r_{a_i}^{F_i}= r_{a_{i+1}}^{F_i}$ and by the preceding paragraph $r_{a_{i+1}}^{F_i}= r_{a_{i+1}}^{F_{i+1}}$ for all $i$. It follows that $r_a^{F_1}= r_b^{F_m}$. Of course, the corresponding result for blue vertices 
holds if $a$ and $b$ are connected by a red path. 

Conversely, it is clear that if $a$ and $b$ are different vertices belonging to 2-cells $F'$ and $F''$, respectively, and $r_a^{F'}= r_b^{F''}$, then $a$ and $b$ are connected by a blue path. Indeed, there exists a sequence of vertices $b_1,\dots, b_m$ and corresponding 2-cells $F_1,\dots, F_m$ such that $a=b_1, F'=F_1$ and $b=b_m, F''=F_m$ and $r_{b_i}^{F_i}$ is identified with $r_{b_{i+1}}^{F_{i+1}}$  
either through a gluing of $\tau_{F_i}$ to $\tau_{F_{i+1}}$ along an edge containing $b_i$, in which case $b_i=b_{i+1}$, or else $F_i=F_{i+1}$ and $b_i$ is connected to $b_{i+1}$ by a blue edge in $F_i$. Similarly, if $b_a^{F'}= b_b^{F''}$ then $a$ and $b$ are connected by a red path.

This shows that the identifications pertaining to $K_S$ and $K'_S$ are the same and hence $K_S=K'_S$ as claimed. We are now ready to prove Lemma \ref{le7}.

\medskip
 \noindent
{\it Proof of Lemma \ref{le7}.} 
Let $S_1,S_2,\ldots  ,S_N =S $ be a local construction (see \cite{dj,bz}) of the midsection $S$.  This means that $S_1$ is a coloured 
2-cell and $S_{n+1}$ is obtained from $S_n$ by either (i) gluing a coloured 2-cell to $S_n$ along an edge $e$ in $\pa S_n$ or (ii) 
by identifying two edges $e_1$ and $e_2$ 
in $\pa S_n$ which have the same colour and share a vertex. The existence of such a construction is well known and easy to establish
in the 2-dimensional case. For further details we refer to \cite{bz,dur}.  Evidently, all the $S_n$'s have the topology of a disc and the 2-cells as well as interior edges of $S_n$ can be identified with corresponding 2-cells and interior edges in $S$. 

The correspondence between tetrahedra $\tau_F$ in $K_S$ and 2-cells $F$ of $S$ and the local construction of $S$ gives rise to a sequence of coloured simplicial complexes $K_n$, $n=1,\ldots ,N$, where $K_n$ is defined by gluing the tetrahedra $\tau_F$ assigned to the 2-cells in $S_n$ along the interior edges of $S_n$. In particular, $K_1$ is a single tetrahedron and $K_N=K_S$, since $S_N=S$ and $K_S=K'_S$ as shown above. Moreover, $K_{n+1}$ is obtained from $K_n$ by gluing a tetrahedron $\tau_F$ to a tetrahedron $\tau_{F'}$ in $K_n$ along an edge $\br ab\kt$. In case $\tau_F$ is not in $K_n$ already it is clear that the topological class of $K_n$ equals that of $K_{n+1}$. We shall now argue that the same holds if $\tau_F$ and $\tau_{F'}$ both belong to $K_n$. Since $K_1$ is homeomorphic to the 3-ball the same will consequently hold for $K_S$, and hence the proof of the lemma will be completed.

If $\tau_F$ and $\tau_{F'}$ belong to $K_n$ and are glued along the edge $e$ 
in $K_n$ 
then $F$ and $F'$ belong to $S_n$, and $S_{n+1}$ is obtained from $S_n$ by identifying 
two edges $e_1$ and $e_2$ in $\pa S_n$ sharing a vertex $v$, such that the identified edges equal $e$.

\begin{figure}[t]
  \begin{center}
 \includegraphics[width=9truecm]{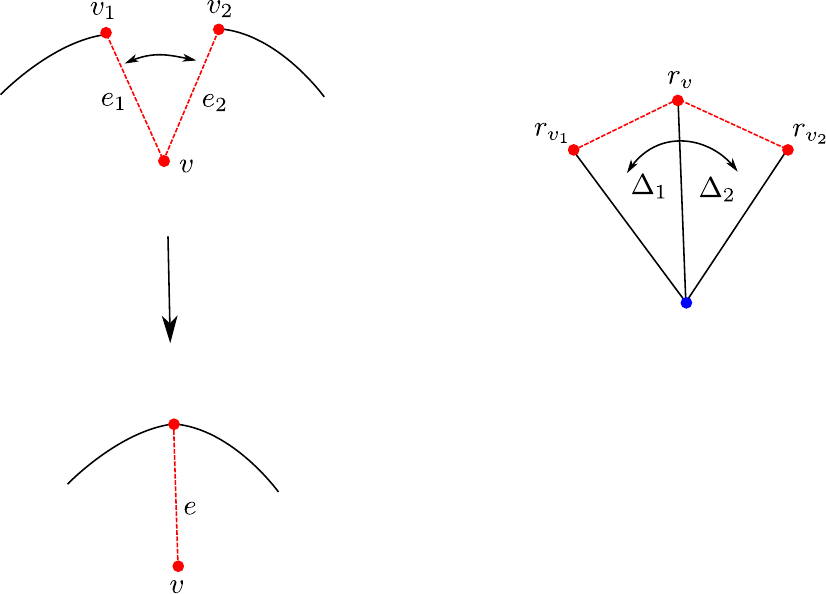}
    \caption{The local construction of the midsection and the corresponding 3-dimensional simplical complex.
    After identifying the edges $e_1$ and $e_2$ we identify the triangles $\Delta_1$ and $\Delta_2$.}
    \label{figY}
      \end{center}
      \end{figure}

Let us set $e_1=\br vv_1\kt$ and $e_2=\br vv_2\kt$ and assume $e$ is red.  
Then $K_{n+1}$ is obtained from $K_n$ by identifying the 
triangles $\Delta_1=(r_{v_1}^{F_1}r_v^{F_1}b_v^{F_1})$ and $\Delta_2=(r_{v_2}^{F_2}r_v^{F_2}b_v^{F_2})$ which share the edge $(r_v^{F_1}b_v^{F_1}) = (r_v^{F_2}b_v^{F_2})$ in $K_n$.
This is illustrated on Fig.~\ref{figY}.
We need to show that no further identifications of simplices are implied.

Additional identifications can only arise if there is a vertex $x\in\pa K_n$ which is a neighbour
of both $r_{v_1}^{F_1}$ and $r_{v_2}^{F_2}$ 
in which case the edges $(xr_{v_1}^{F_1})$ and $(xr_{v_2}^{F_2})$ are identified in the step from $K_n$ to $K_{n+1}$.
We claim that no such $x$ exists except $b_v^{F_1}$ and $r_v^{F_1}$.

\begin{figure}[h]
  \begin{center}
 \includegraphics[width=8truecm]{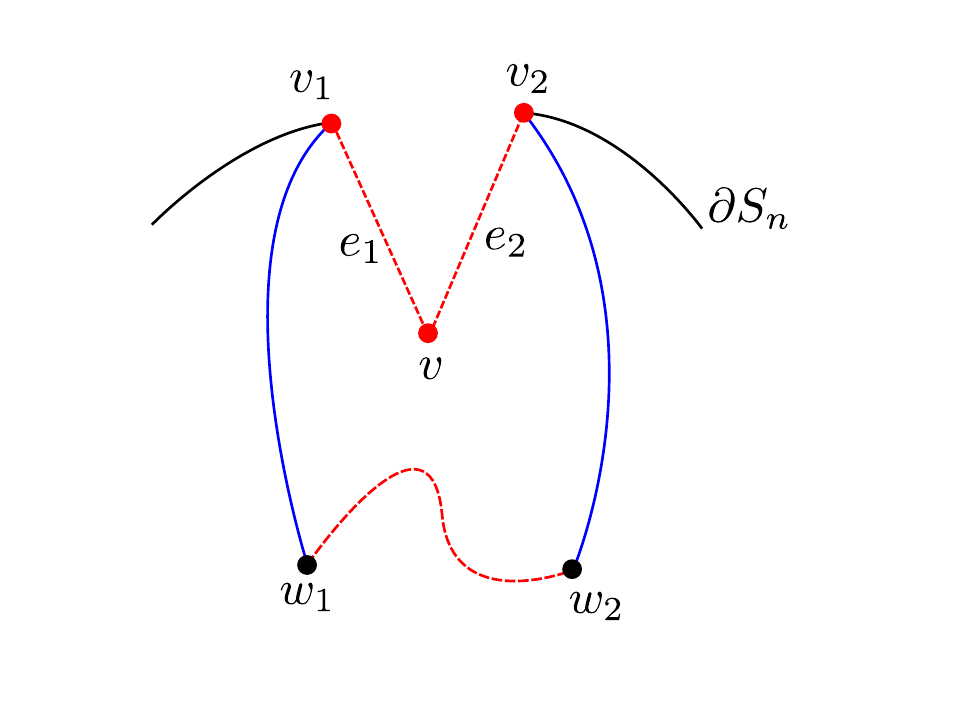}
    \caption{Paths arising in the local construction of the midsection when $x$ is blue.}
    \label{Z0}
      \end{center}
      \end{figure}

First, suppose $x$ is a blue vertex, $x\neq b_v^{F_1}$.  Then 
there are two vertices $w_1$ and $w_2$ in the midsection such that $b_{w_1}^{F'_1}=x=b_{w_2}^{F'_2}$ 
and $r_{w_i}^{F'_i}=r_{v_i}^{F_i}, i=1,2$, in $K_n$ for some 2-cells $F'_1, F'_2$ in $S_n$. This implies that $w_1$ and $w_2$ are connected by a red path in $S$ while $w_i$ is connected to $v_i$ by a blue path 
in $S$ for $i=1,2$.  When we now take the step 
from $S_n$ to $S_{n+1}$ and identify $e_1$ and $e_2$ the vertices $v_1$ and $v_2$ get identified and 
hence $w_1$ and $w_2$ are connected both by a blue path in $S$  
and by a red path, see Fig.\  \ref{Z0}.  
Hence, $w_1=w_2$, and we get a closed blue path enclosing  the red edge $e$ which is impossible by 
$(\beta_1)$.  

\begin{figure}[h]
  \begin{center}
 \includegraphics[width=6truecm]{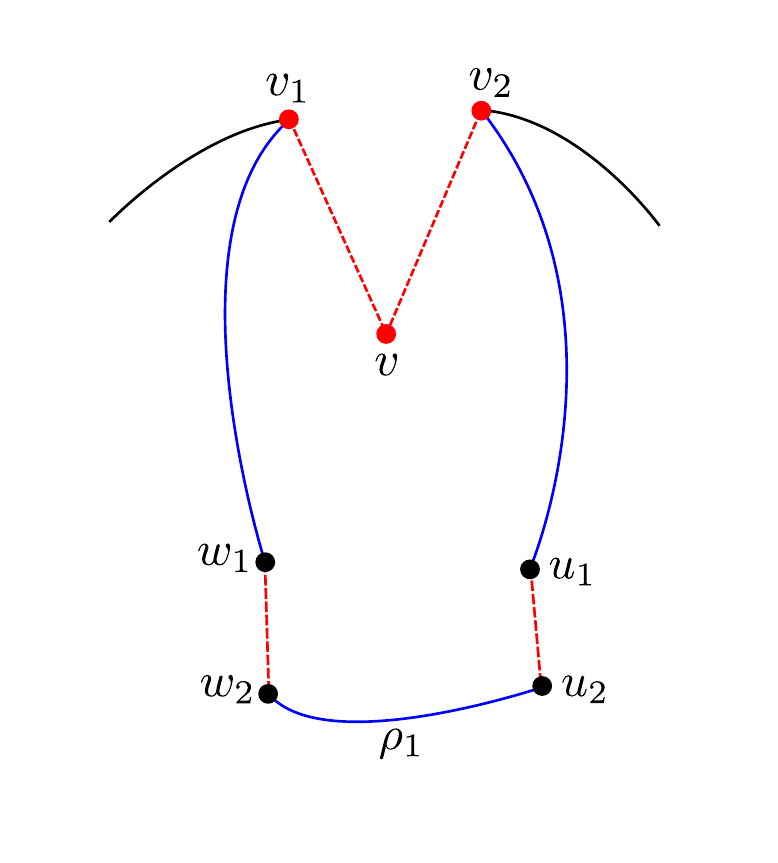}
    \caption{Paths arising in the local construction of the midsection when $x$ is red.}
    \label{Za}
      \end{center}
      \end{figure}

Now suppose $x$ is a red vertex.  Then there are four vertices $w_1,w_2,u_1,u_2$ in the midsection such that 
$\br w_1w_2\kt$ and $\br u_1u_2\kt$ are red edges and 
$x=r_{w_2}^{F'}=r_{u_2}^{F''}$, $r_{w_1}^{F'}=r_{v_1}^{F_1}$, $r_{u_1}^{F''}=r_{v_2}^{F_2}$ for some 2-cells $F', F''$ in $S_n$. It follows that $w_2$ and $u_2$ are connected by a blue path $\rho_1$ and similarly that  
$w_1$ is connected to $v_1$ and $u_1$ is connected to $v_2$ by a blue path in $S$, see Fig.\ \ref{Za}. 

\begin{figure}[h]
  \begin{center}
 \includegraphics[width=10truecm]{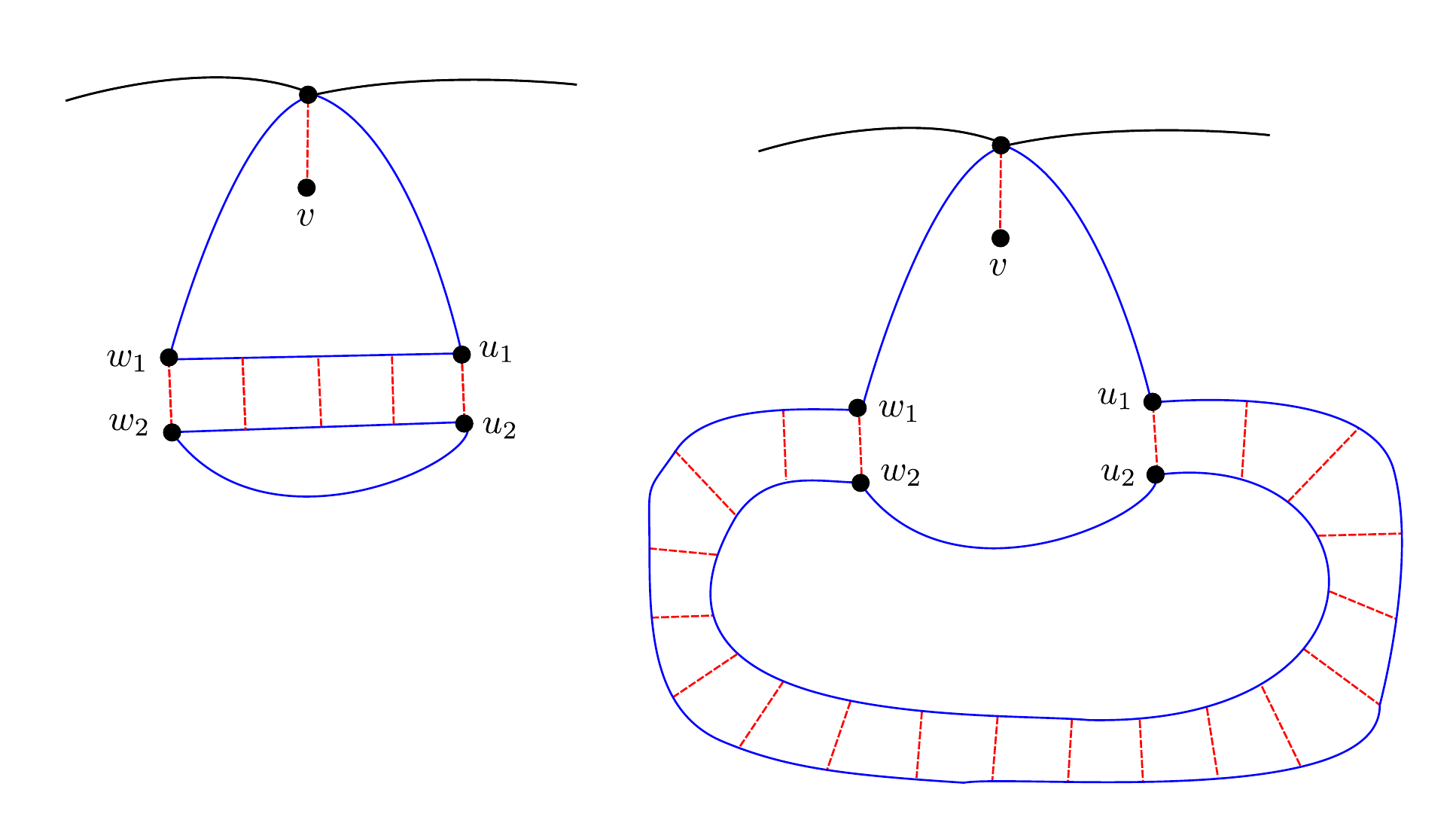}
    \caption{Paths of quadrangles arising in the local construction of the midsection when $x$ is red.}
    \label{figZ}
      \end{center}
      \end{figure}

Upon merging $e_1$ and $e_2$ we get as above a blue path $\rho_2$ from $w_1$ to  $u_1$.
It follows by condition $(\gamma)$ 
that the edges $\br w_1w_2\kt$ and $\br u_1u_2\kt$ are connected by a blue path of quadrangles.
This path of quadrangles either lies inside the closed loop made up of $\rho_1, \rho_2$ and the edges $\br w_1w_2\kt$ and $\br u_1u_2\kt$ or outside it, see Fig.\ \ref{figZ}.
In both cases we obtain a closed blue loop containing a red edge in its interior which violates $(\beta_1)$.
This completes the proof of Lemma~\ref{le7}.
\hfill $\Box $

\section{Sphere-slices}

In this section we generalise Theorem \ref{thm1} to the case of sphere-slices.   
It is clear from the proofs of Lemmas \ref{le2} and \ref{le4} that the midsection of any sphere-slice has 
properties $(\alpha)$ and $(\gamma)$, while ($\beta_2$) is irrelevant as it refers to the boundary. Furthermore, the proof of 
Lemma \ref{le3}  shows that an analogue of ($\beta_1$) still holds in the following form:
\begin{itemize}
\item[($\beta$)] Each simple closed red (or blue) path divides the midsection into two components,
one of which is red (blue).
\end{itemize}
Since a simplicial decomposition of the 2-sphere requires at least 4 triangles it follows that $S$ must contain at least 4 triangles of each colour. 
Let $\phi$ denote the mapping that takes any sphere slice $K$ to its midsection $S$.

\begin{definition}\label{def5}
We let $\cS$ denote the set of all coloured cell complexes $S$ homeomorphic to the 2-sphere (with cells as described previously) which have at least four triangles of each colour and satisfy the conditions 
($\alpha$), ($\beta$), and ($\gamma$).

\end{definition}
\noindent
\begin{lemma}  \label{le8}
Property ($\delta$) holds for all $S\in \cS$.

\end{lemma}
\noindent
{\it Proof.}  
Applying the same argument as in the proof of Lemma \ref{le6} we conclude that if ($\alpha$), ($\beta$), and 
$(\gamma)$ hold for a coloured spherical cell complex $S$ and there are two different red triangles in
 $S$ whose vertices are pairwise connected by blue paths, then there exist three paths of quadrangles connecting the edges of the triangles pairwise and the complement of those paths consists entirely of blue triangles together with the two given red triangles in $S$. In particular, there can be only two red triangles, which contradicts $S\in\cS$. Clearly, a similar argument holds for blue triangles.   
\hfill $\Box $

We now state and prove the main result in this section.

\noindent
\begin{theorem} \label{thm2}
  The map $\phi: K\to S$ is a bijection from the set $\cC\cS$ of causal sphere-slices to the 
  set $\cS$ of coloured 2-dimensional cell complexes.

\end{theorem}

\noindent
{\it Proof.}
Given $S_0\in\cS$ it is enough to construct a sphere-slice $K$ such that $\phi (K)=S_0$.  The idea of the proof 
is to cut a piece out of $S_0$ in such a way that we obtain a cell complex $S$ in $\cC\D$.  Then we use 
Theorem \ref{thm1} to obtain a disc-slice with the given midsection $S$.  Finally, we fill
in the "hole`` of the disc-slice to obtain a sphere-slice with midsection $S_0$.   We divide the argument into 5 parts.
\medskip

\begin{figure}[h]
  \begin{center}
 \includegraphics[width=8truecm]{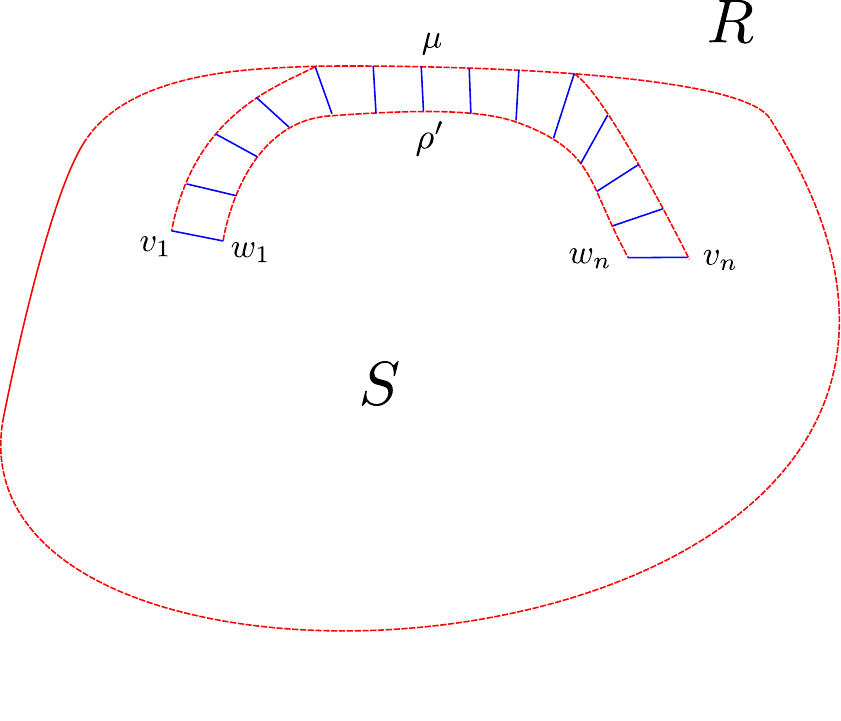}
    \caption{A maximal path of quadrangles in $S$ touching the boundary.}
    \label{Fig18}
      \end{center}
      \end{figure}

\begin{figure}[h]
  \begin{center}
 \includegraphics[width=10truecm]{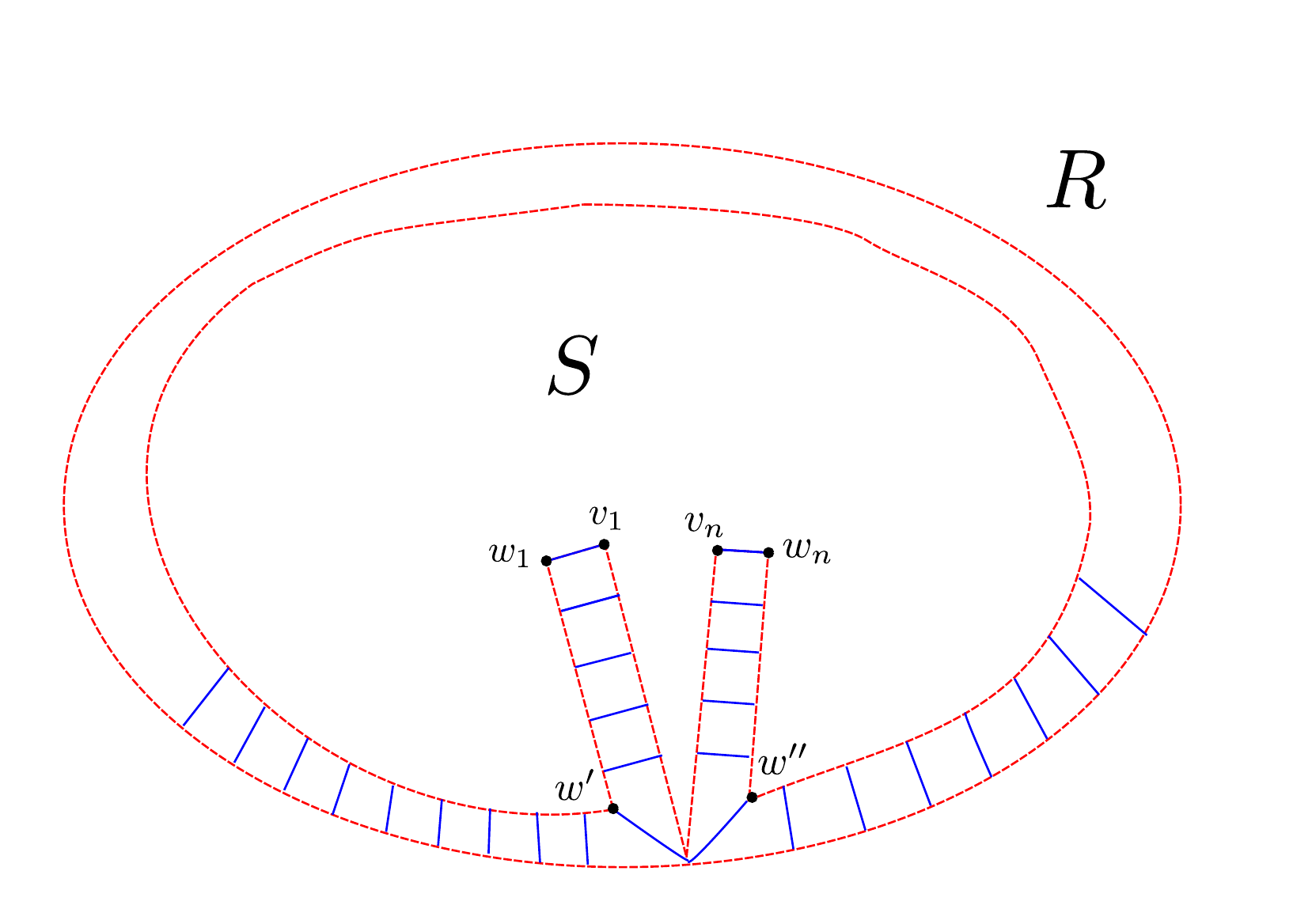}
    \caption{If $\pa S=\mu$ then $w'$ and $w''$ are connected both by a red and by a blue path.}
    \label{Fig20}
      \end{center}
      \end{figure}

\noindent
(1) Let $R$ be a maximal edge-connected cluster of red triangles in $S_0$ and let $S=S_0\setminus {\rm int}(R)$.  
Condition $(\beta)$ and the Jordan Curve Theorem imply that $R$ and $S$ are 
(closed) discs.   Moreover,
since $R$ is maximal, all edges in $\pa S$ are red and contained in quadrangles in $S$.  Pick one of these 
quadrangles $q$ and let $\Gamma =(q_1,q_2,\ldots ,q_n)$ be the maximal blue path of quadrangles containing $q$.
Let $\rho=(v_1,v_2,\ldots ,v_n)$ be the red path in $\Gamma$ that shares at least one red edge with $\pa S$,
see Fig.\ \ref{Fig18}.  
      The other red path in $\Gamma$, $\rho'=(w_1,w_2,\ldots ,w_n)$, 
  shares no vertex with $\pa S$ as a consequence of condition $(\alpha)$. Likewise, $\rho$ and $\rho'$ do not intersect and neither $\rho$ nor $\rho'$ can have multiple vertices, since it would contradict 
	$(\alpha)$. Moreover, if $1\leq i<j\leq n$ are such that $v_i,v_j\in \pa S$, then $v_k\in\pa S$ for all $k$ between $i$ and $j$ since otherwise $R$ is not maximal due to $(\beta)$.  Hence, $\Gamma$ intersects $\pa S$ along a curve segment $\mu$, see Fig.\ \ref{Fig18}. 
Note that $\mu$ does not contain all of $\pa S$ because this would again contradict condition 
$(\alpha)$, see Fig.\ \ref{Fig20}.
     
\begin{figure}[h]
  \begin{center}
\includegraphics[width=10truecm]{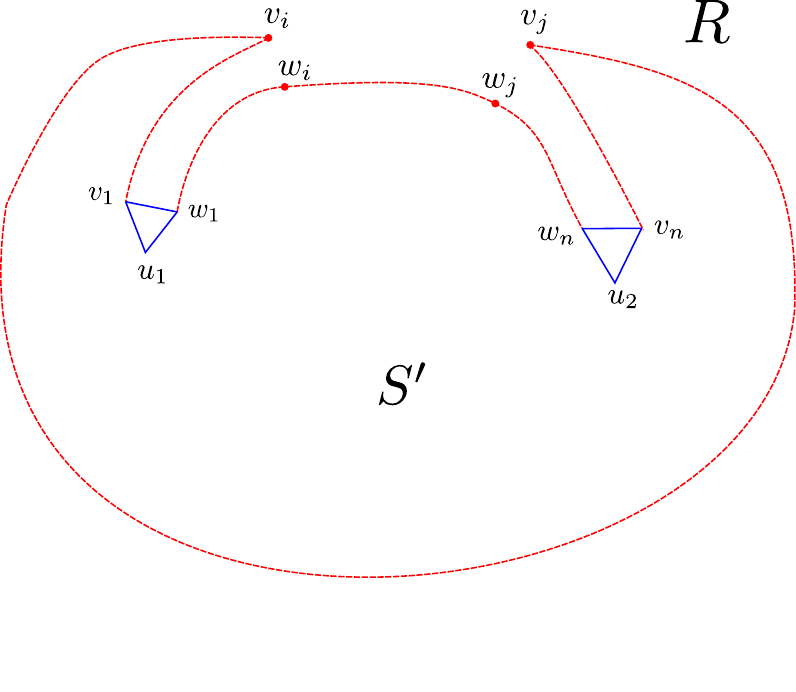}
    \caption{The cell complex $S'$.}
    \label{Fig21}
      \end{center}
      \end{figure}

The above observations show that the boundary of the strip covered by the path of quadrangles 
$\Gamma$, which consists 
of $\rho$ and $\rho'$ and the two blue edges $\br v_1w_1\kt$ and $\br v_nw_n\kt$, is a simple curve.
This curve shares exactly a segment $\mu=(v_i,v_{i+1},\ldots ,v_j)$ with $\pa S$ where
$1\leq i<j\leq n$.  Hence, removing the strip covered by $\Gamma$ from $S$ we obtain a new disc $S'$ whose boundary consists of the blue edges $\br v_1w_1\kt$ and $\br v_nw_n\kt$ together with the red 
paths $\rho'$, $\pa S\setminus \mu$, $(v_1,\ldots ,v_i)$ and $(v_j,\ldots ,v_n)$, see Fig.\ \ref{Fig21}.
Now consider the blue triangles $\Delta_1$ and $\Delta_2$ in $S'$ containing $\br v_1w_1\kt$ 
and $\br v_nw_n\kt$, respectively.
Denote the third vertex in these triangles by $u_1$ and $u_2$, respectively, see Fig.\ \ref{Fig21}. 
It follows from condition $(\delta)$ (see Lemma \ref{le8}) that $u_1\neq u_2$ and they cannot be connected by a red path.  Moreover, $u_1$ and $u_2$ are not in $\pa S'$ since this would violate $(\alpha)$.  Now, removing $\Delta_1$ and $\Delta_2$ (including edges $\br v_1w_1\kt$ and $\br v_nw_n\kt$) from $S'$
we obtain another 2-dimensional cell complex $S''$ 
with the topology of a disc and a boundary consisting of the blue paths
$(v_1,u_1,w_1)$ and $(v_n,u_2,w_n)$ and the same red paths as $\pa S'$.

\medskip
\noindent
(2) We now claim that $S''$ belongs to $\cS\cD$.  Since paths in $S''$ are also paths in $S_0$, conditions $(\alpha)$ and $(\beta_1)$ are obviously fulfilled.  Furthermore, $S''$ contains at least two blue triangles, and since there exist red edges in $\pa S''$ the existence of at least one red triangle in $S''$ will follow once $(\beta_2)$ has been established for $S''$.

In order to verify $(\beta_2)$ we recall that $u_1$ and $u_2$ cannot be connected by a red path.  Connecting $u_1$ to $w_n$ or $v_n$ or connecting $u_2$ to $w_1$ or $v_1$ by a red path evidently contradicts $(\alpha)$. 
Similarly, a red path from $v_1$ to $w_n$ or from $w_1$ to $v_n$ contradicts $(\alpha)$. 
 This proves $(\beta_2)$ for $S''$ as far as red paths are concerned. 

Now let $\eta$ be a blue path in $S''$ connecting the vertex $w_k$ to a vertex $v$ in the other red arc of $\pa S''$.  Viewing  $\eta$ as a path in $S_0$ and extending it by the edge $\br w_kv_k\kt$ we obtain a blue path $\eta'$ in $S_0$ connecting $v_k$ to $v$.
If $v\neq v_k$ then condition $(\alpha)$ is violated.   If $v=v_k$ then $\eta'$ is a closed blue path in $S_0$ with edges of both colours inside and outside unless $k=1$ or $k=n$.  This proves condition $(\beta_2)$.

In order to verify condition $(\gamma)$ for $S''$ let $e_1=\br x_1y_1\kt$ and $e_2=\br x_2y_2\kt$ be two blue edges in $S''$ such that $x_1$ and $x_2$ as well as $y_1$ and $y_2$ are connected by red paths.  Then, since $(\gamma)$ holds for $S_0$, there is a blue path of quadrangles $\Lambda$ in $S_0$ connecting $e_1$ 
and $e_2$.  Since the blue boundary edges of $S''$ are not contained in quadrangles outside $S''$ the path $\Lambda$ cannot leave $S''$.  This establishes 
condition $(\gamma)$ for the blue edges. 

\begin{figure}[h]
  \begin{center}
\includegraphics[width=10truecm]{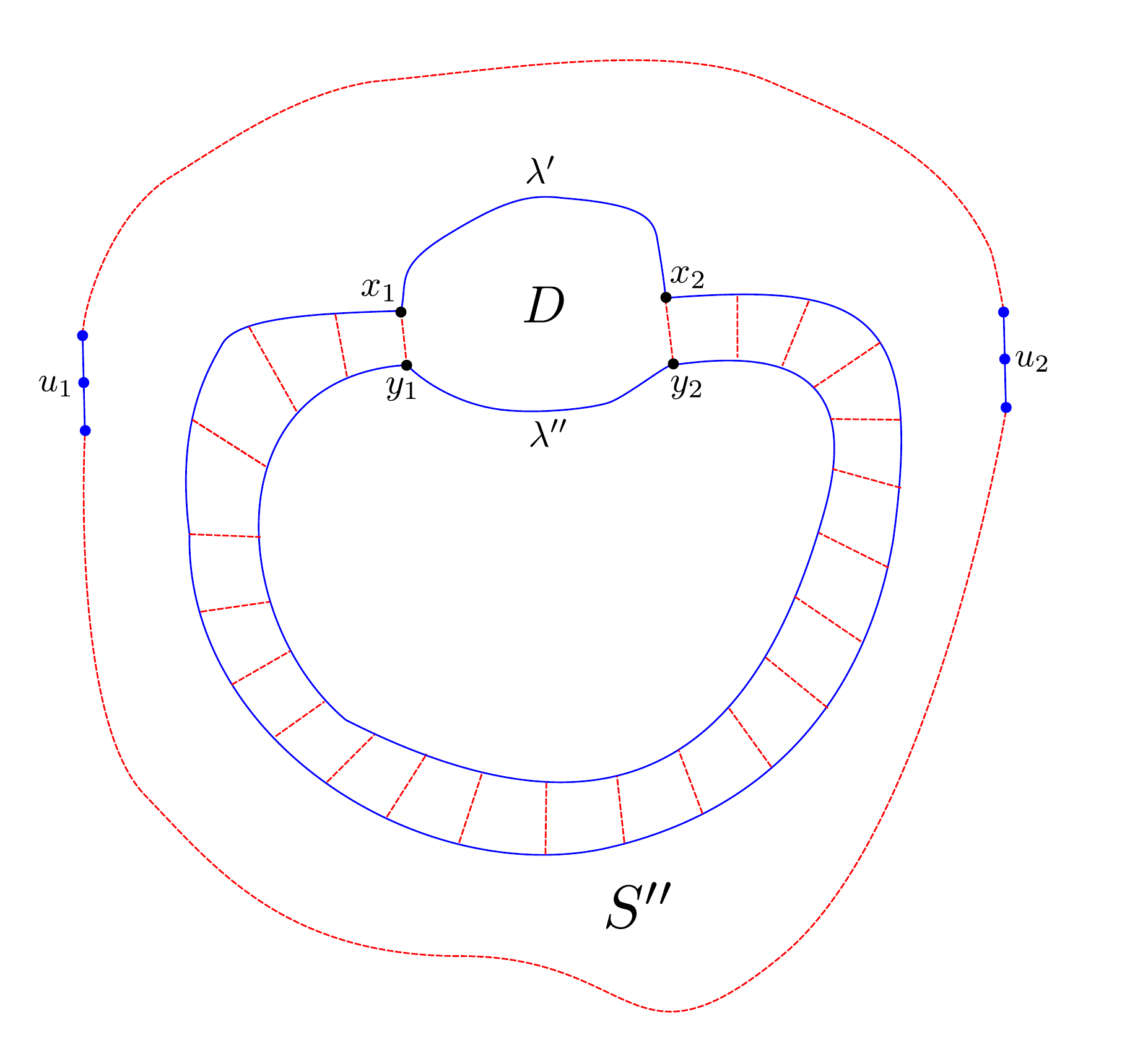}
    \caption{A path of quadrangles outside $D$ connecting the edges $e_1$ and $e_2$
    and the blue paths $\lambda'$ and $\lambda''$ connecting their endpoints. A priori, parts of the path may be outside $S''$.}
    \label{Fig22}
      \end{center}
      \end{figure}

Let now $e_1=\br x_1y_1\kt$ and $e_2=\br x_2y_2\kt$ be two red edges in $S''$ such that $x_1$ and $x_2$ as well as $y_1$ and $y_2$ are connected by blue paths $\lambda'$ and $\lambda''$.  By condition 
$(\alpha)$ the paths $\lambda'$ and $\lambda''$ do not intersect and we can assume without loss of generality that they have no multiple vertices.
Hence, the closed curve consisting of $\lambda'$, $\lambda''$ and the two edges $e_1$ and $e_2$ is simple and encloses a disc $D$ in $S''$.   Since condition $(\gamma)$ holds for $S_0$ there exists a red path $\Gamma'$ of quadrangles in $S_0$ 
connecting $e_1$ and $e_2$ and evidently $\Gamma'$ is either contained in $D$ or in its exterior.  However, the latter case can be excluded as follows. Observe that neither $D$ nor $\Gamma'$ can intersect $R$. It follows that one of the closed curves made up of either $\lambda'$ or 
$\lambda''$ and one of the blue paths in $\Gamma'$ must separate $R$ (in $S_0$) from the union of $D$ and $\Gamma'$ in contradiction with condition $(\beta)$, see Fig.\ \ref{Fig22}. Note, however, that the curve in question is not necessarily simple, since $\lambda'$ may touch the blue path in $\Gamma'$ with the same endpoints, and similarly for $\lambda''$. Using property ($\beta$) one can in this case remove closed parts of the curve so as to obtain a simple closed blue curve with the same separating property.  
Thus $\Gamma'$ is contained in $D$ (and hence in $S''$) and this proves that condition $(\gamma)$ holds for $S''$.  

\medskip
\noindent
(3) We can now apply Theorem \ref{thm1} to conclude that there is a unique disc-slice $K''\in\cC \cD$ 
whose midsection is $S''$.
In view of the structure of $\pa S''$, the side of $K''$ 
consists of two pairs of adjacent backwards directed triangles and two arrays of forward directed triangles 
corresponding to the two red arcs in $\pa S''$.  

\begin{figure}[h]
  \begin{center}
\includegraphics[width=8truecm]{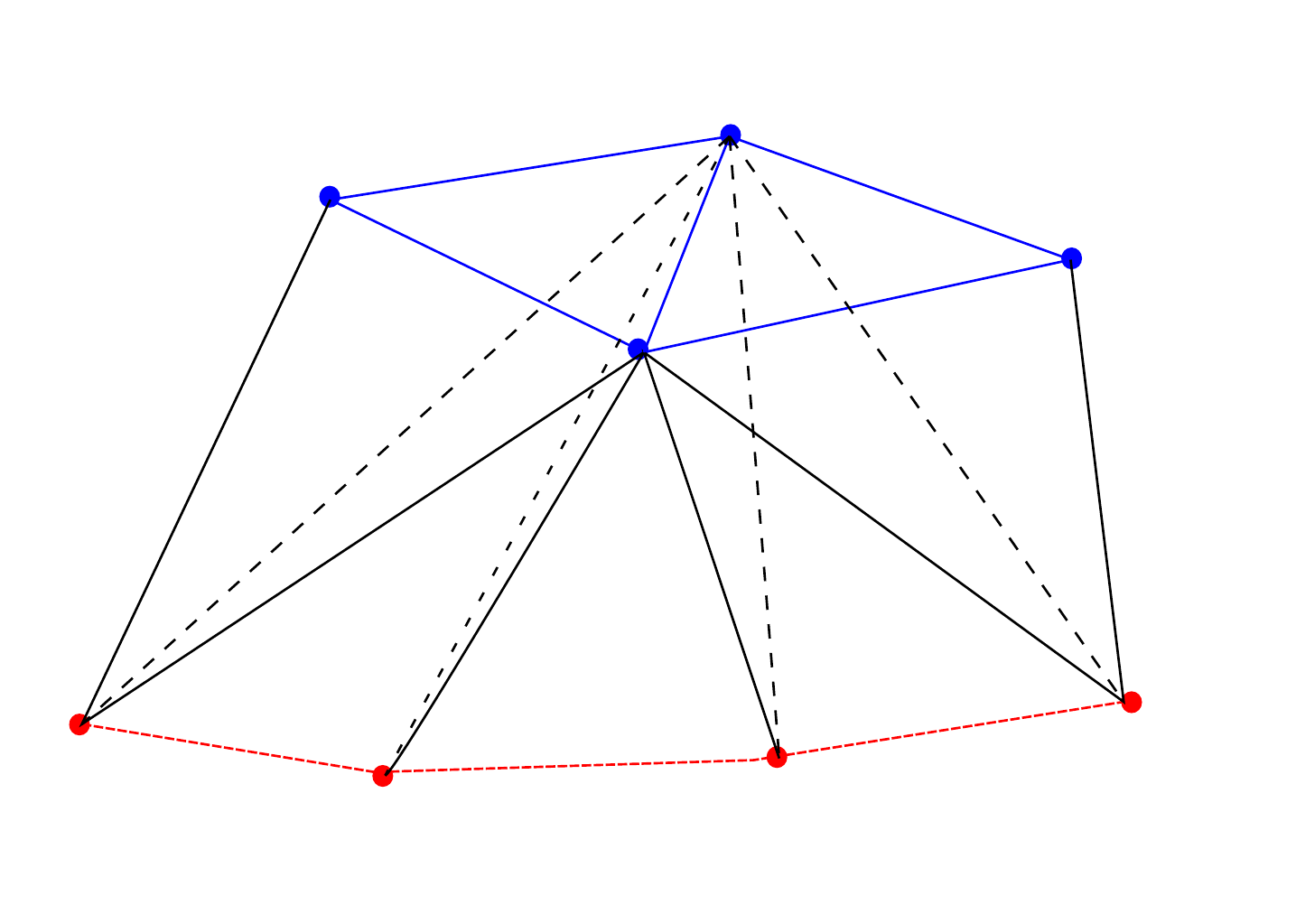}
    \caption{The cell complex $K_0$ in the case $n=3$.}
    \label{Fig23}
      \end{center}
      \end{figure}

We now first construct from $K''$ a new simplicial complex $C$ by gluing to $K''$ the tetrahedra corresponding to the 
quadrangles $q_1,\ldots ,q_n$ in $\Gamma$ and those corresponding to the blue triangles at the ends of $\Gamma$.
One way to accomplish this is to consider the simplicial complex $K_0$ whose midsection is $\Gamma$ with the two 
blue triangles attached to its ends.  Clearly $K_0$ is a ball and we may think of it as a causal slice whose side consists of two pairs of adjacent backwards directed triangles and two arrays of forward directed triangles of size $n$ whose red edges are pairwise
 identified and constitute $D_{\rm red}$. On the other hand, $D_{\rm blue}$ consists of two triangles with one common edge, see Fig.\ \ref{Fig23}.  
 Now consider one array of forward directed triangles in $\pa K_0$ together with the two adjacent backwards directed triangles.  These triangles 
 form a disc in $\pa K_0$ and we can glue $K_0$ to $K''$ along this disc and the corresponding disc in $\pa K''$ or, more precisely, the gluing is performed along the edges $\br u_1w_1\kt, \br w_1w_2\kt,\dots,\br w_nu_2\kt$,  
see Fig.\ \ref{Fig21}.   Since both $K_0$ and
 $K''$ are simplicial balls, so is the resulting complex $K''\# K_0$.  Its Euler characteristic is therefore $-1$. 
 
 Note also that $K''\# K_0$ has a side consisting of two pairs of adjacent backwards directed triangles 
 and two arrays of forward directed triangles, one of which has length $n$.  The blue edges in the backwards directed triangles form a closed curve.  We now identify the two backwards directed triangles in each pair, i.e.\ we glue tetrahedra in $K''\# K_0$ along the edges $\br u_1v_1\kt$ and $\br u_2v_n\kt$.  Doing this, the number of triangles in the simplicial complex decreases by 2, the number of edges decreases by 4 and the number of vertices decreases by 1 and we obtain a new simplicial complex $C'$ with Euler characteristic $-2$.
 
 The side of $C'$ is a closed circuit of forward directed triangles sharing a single blue vertex.  We now obtain 
 the simplicial complex $C$ by gluing tetrahedra in $C'$ along the egdes $\br v_kv_{k+1}\kt$  
 in the midsection, for $k=1,\ldots ,i-1$ and 
 $k=j,\ldots ,n-1$.  It is easy to check that the Euler characteristic does not change with these identifications so
 $$
 \chi (C)=-2.
 $$
 By construction, the midsection of $C$ is $S_0\setminus R$.
 
 \medskip
 \noindent
 (4)  Now return to the disc $R$ that was removed from $S_0$ in the first step.  Let $\overline{R}$ be the cone over $R$, i.e.\ 
 $\overline{R}$ consists of the tetrahedra obtained by adding a common blue 
 vertex to all the triangles in $R$ or alternatively, by gluing the tetrahedra corresponding to 
triangles in $R$ along interior edges in $R$.  We can now 
 glue $\overline{R}$ to $C$ along their sides, i.e.\ along the edges in $\pa S =\pa R$.  The 
 resulting simplicial complex $K$ has Euler characteristic
 $$
 \chi (K)=\chi (C)+\chi(\overline{R})+\chi_2 (\overline{\partial R})=-2\,,
 $$
 since $\chi (\overline{R})=-1$, and $\chi_2(\overline{\partial R}) = 1$ is the Euler characteristic of the cone $\overline{\partial R}$ over $\pa R$, which is in fact a disc.
 
 \medskip
 \noindent
 (5) Since $K''$ is a disc-slice it is evident from the preceding construction that $\pa K$ consists of two triangulated
 2-spheres, one red and one blue.  The midsection of $K$ is by construction $S_0$.
By gluing on cones over the two boundary components of $K$ we obtain a simplicial complex $\tilde{K}$ 
with Euler characteristic 0.  Since $\tilde{K}$ is clearly a pseudomanifold we conclude 
(see \cite{seifert} p.\ 216) that $\tilde{K}$ is also a manifold.  In fact, $\tilde{K}$ is simply 
connected since any closed curve can be deformed to a closed curve on the midsection which is simply connected. 
Hence, $\tilde{K}$ is a 3-sphere by \cite{perel} and $K$ is a cylinder as desired.\footnote{This can presumably 
be proven by lesser means  but the argument does not seem to be entirely trivial and we will not elaborate on it here.}
This completes the proof of Theorem \ref{thm2}.
\hfill $\Box $

\section{Discussion}\label{sec:6}

A few remarks pertaining to extensions and variations of the present work are in order. As already indicated, the requirement in Definition~\ref{def2} that $D_{\rm red}$ and $D_{\rm blue}$ be discs is unnecessarily restrictive. 
Thus, replacing $(ii)$ of Definition~\ref{def2} by  

\medskip

{\it (ii') all monocoloured simplices of $K$ belong to the boundary $\partial K$, such that the red (resp.\ blue) ones form a connected and simply connected subsimplex $D_{\rm red}$ (resp.\ $D_{\rm blue}$) of $\partial K$,}

\medskip

\noindent leads to a convenient class of causal triangulations. Let us call this class $\cC\cD'$. In particular, any coloured tetrahedron belongs to $\cC\cD'$ (but not to $\cC\cD$) and so does the complex depicted in Fig.\ \ref{Fig23}. 

By inspection of the proofs given above it is seen that condition ($\beta_2$) is used in the construction of a causal disc slice from a cell complex in $\cC\cD$ solely to ensure that $\partial D_{\rm red}$ and $\partial D_{\rm blue}$ are simple curves. It follows that dropping ($\beta_2$) one obtains a one-to-one correspondence between $\cC\cD'$ and the set $\cS\cD'$ of coloured cell complexes homeomorphic to a disc and fulfilling conditions ($\alpha$), ($\beta_1$), and ($\gamma$). 

On a different note, one may observe that condition ($\gamma$) is used partly to ensure the validity of condition ($\delta$) (see Lemma~\ref{le6}) and otherwise only in the final part of the proof of Lemma~\ref{le7}
ensuring that the complex $K$ is a simplicial ball. It is natural to consider replacing property ($\gamma$) by property ($\delta$) or dropping both of them. This would require stepping outside the category of simplicial complexes and adopt a different setting encompassing singular triangulations. 

Finally, one might also envisage characterising causal triangulations in higher dimensions, which evidently would lead to more involved higher dimensional coloured cell complexes of which little is known at present.   In \cite{DJ} it is explained how the midsections of 
4-dimensional causal triangulations are made up of tetrahedra and prisms which replace the triangles and quadrangles which make
up the midsections considered in this paper.

\medskip
\noindent
{\bf Acknowledgements.}  BD acknowledges support from the Villum Foundation via the QMATH Centre of Excellence 
(Grant no. 10059).

\end{document}